\documentclass[conference]{IEEEtran}

\AtBeginDocument{%
  }

\ifCLASSINFOpdf
\else
\fi

\usepackage{hyperref}
\usepackage{url}
\usepackage{tikz}
\usepackage{amsmath}
\usepackage{array}
\usepackage{booktabs}
\usepackage{multirow}
\usepackage{multicol}
\usepackage{makecell}
\usepackage{textcomp}
\usepackage{amssymb}
\usepackage{amsfonts}
\usepackage{wasysym}
\usepackage{graphicx}
\usepackage{pifont}
\usepackage{bm}
\usepackage{xcolor}

\newcommand{\OutAttenMul}{\textit{OutAttnMult}}
\newcommand{\OutSoftMax}{\textit{OutSoftMax}}
\newcommand{\UVerify}{\textit{U-Verify}}
\newcommand{\name}{$\mathrm{TwinShield}$}

\usepackage{amsthm}
\begin{document}

\title{Securing Transformer-based AI Execution via Unified TEEs and Crypto-protected Accelerators}

\author{Jiaqi Xue,
  Yifei Zhao,
  Mengxin Zheng,
  Fan Yao,
  Yan Solihin,
Qian Lou\\
University of Central Florida 
 
}

\maketitle

\begin{abstract}

Recent advances in Transformer models, e.g., large language models (LLMs), have brought tremendous breakthroughs in various artificial intelligence (AI) tasks, leading to their wide applications in many security-critical domains. Due to their unprecedented scale and prohibitively high development cost, these models have become highly valuable intellectual property for AI stakeholders and are increasingly deployed via machine learning as a service (MLaaS). However, MLaaS often runs on untrusted cloud infrastructure, exposing data and models to potential breaches. Mainstream protection mechanisms leverage trusted execution environments (TEEs) where confidentiality and integrity for secretive data are shielded using hardware-based encryption and integrity checking. Unfortunately, running model inference entirely within TEEs is subject to non-trivial slowdown, which is further exacerbated in LLMs due to the substantial computation and memory footprint involved. Recent studies reveal that the hybrid TEE-based scheme offloading partial model inference operations to the untrusted accelerators (e.g., GPU) is a promising solution. 
However, prior offloading schemes fail to ensure dual protection of data and model in Transformer inference, as they cannot securely offload critical operations, i.e., \texttt{Attention} and \texttt{SoftMax}, forcing these computations to remain confined within TEEs. To address these challenges, we propose \name{}, a framework enabling secure Transformer inference in heterogeneous TEE and accelerator systems with dual protection for both model and data. \name{} offloads $\sim 87\%$ of computation to GPUs and delivers $4.0\times$ to $6.1\times$ speedups over previous approaches across various Transformer models.

\end{abstract}

\IEEEpeerreviewmaketitle

\section{Introduction}
Transformers~\cite{vaswani2017attention} have demonstrated outstanding performance on a wide range of domains including computer vision~\cite{dosovitskiy2020image} and natural language processing~\cite{devlin2018bert, touvron2023llama}, which are the building blocks for many emerging applications such as chat-bots~\cite{floridi2020gpt} and medical image analysis~\cite{myszczynska2020medical1, richens2020medical2}. As Transformers become increasingly popular, the \textit{confidentiality} and \textit{integrity} of the inference services become a critical concern, especially in confidentiality-sensitive sectors such as healthcare~\cite{myszczynska2020medical1, richens2020medical2}, finance~\cite{heaton2017finance}, and personal assistant applications~\cite{mclean2019apphome, foerster2016lappcom}. 
Given the substantial size and deployment complexity of these models, cloud-based Transformer-as-a-Service (TaaS) has emerged as a widely adopted solution for end users to access these state-of-the-art models in a cost-efficient way~\cite{natarajan2023chex,zhangcipherprune, zheng2023primer}. 




In these services, data provided by clients, such as personal health information (including sleep patterns, pulse, and heart rate) and banking records, are highly private. 
However, it is widely known that remote computation (as in the cloud) may not be trusted as adversaries can exploit either privileged system software~\cite{pahima2022attck1} or hardware vulnerabilities~\cite{tung2021attack2} to compromise data privacy and integrity. This becomes particularly worrisome for Transformer-based systems where exposure of sensitive data can result in tremendous breaches of personal privacy (e.g., for clients). 
Moreover, an adversary may maliciously tamper with the model and its computation, leading to severe integrity compromise that introduces catastrophic system consequences. In summary, offering data confidentiality and inference integrity is imperative.

Trusted Execution Environments (TEEs), such as Intel SGX~\cite{sgx,sgx_attestation}, offer an environment for safeguarding the privacy (and sometimes integrity) for sensitive computation. In systems with TEEs, the CPU is treated as the root of trust.  The processor shields individual secure domains (i.e., enclaves) from privileged system software attacks via hardware-enforced isolation. Furthermore, counter-mode encryption and integrity tree-based data verification are performed by the TEE-hardware to protect the breach and tampering of off-chip data belonging to enclaves~\cite{sgx_attestation}.
Accordingly, prior studies have investigated the use of TEEs for secure machine learning inference~\cite{hanzlik2021mlcapsule, tramer2018slalom, sun2023shadownet, shen2022soter}. For instance, MLCapsule~\cite{hanzlik2021mlcapsule} proposes to store CNN models in enclave and perform model inference completely in TEEs, hence protecting computation integrity and the confidentiality of all data.   
Unfortunately, deployment of the entire ML model inside TEEs introduces extremely high overhead due to the limited resources available within TEEs.
Subsequent works~\cite{tramer2018slalom, hashemi2021darknight, sun2023shadownet, shen2022soter, deng2024trinity, yudha2024boostcom} attempt to improve the performance of TEE-based model inference by \textit{outsourcing} certain computation from TEEs to an untrusted external accelerator (e.g., GPUs, FPGAs and ASICs), and \textit{verificating} the computation integrity inside the enclave. While the aforementioned secure ML outsourcing techniques enhance system performance of TEE-only methods, they struggle to outsource sufficient computations to untrusted accelerators from trusted TEEs. The challenges are summarized as follows:



\textit{(I) Confidentiality Challenge: Multiplicative Linear Operations, e.g., Transformer's Attention.} 
We categorize linear operations into two types: additive and multiplicative. Additive operations, such as matrix multiplication involving one \textit{variable} matrix and one \textit{constant} matrix. Multiplicative operations involve matrix multiplication where neither operand is a constant matrix. Prior works show the feasibility of securing execution of the additive operation on untrusted accelerators via secret sharing and Freivalds' algorithm~\cite{freivalds1977probabilistic}, which works well for the traditional convolutional neural networks (CNNs) since the convolution of fixed constant pre-trained kernels and variable inputs is additive operations. However, Transformers include massive multiplicative linear operations where neither operands are constant matrices, rendering prior techniques inapplicable; 

\textit{(II) Confidentiality Challenge: Non-linear Operations, e.g., \texttt{SoftMax}.} We find that different from CNNs where operations in linear layers overwhelmingly dominate the computation for model inference ($>98\%$)~\cite{tramer2018slalom}, non-linear operators in Transformers (i.e., \texttt{SoftMax}) contribute non-trivial computation overhead, especially for inputs with long tokens. 
Therefore, under an outsourcing scheme with linear operation-only offloading, the execution of such non-linear functions in TEEs will become the new bottleneck. As a result, it is necessary to further outsource \texttt{SoftMax} operations for further unleash the performance advantages of off-chip accelerators. Unfortunately, none of the prior mechanisms is able to outsource non-linear operations while maintaining proper model privacy and integrity at the same time. 

\textit{(III) Integrity Challenge: Effective Verification for Multiplicative linear and Non-linear Operations.} Prior works~\cite{tramer2018slalom, hashemi2021darknight, sun2023shadownet} rely on Freivalds' algorithm~\cite{freivalds1977probabilistic} to guarantee the integrity of the outsourced matrix computation, i.e., based on the matrix multiplication's associative law, $A\cdot B = C$ can be verified by $A \cdot (B \cdot r) == C \cdot r$, where $A, B, C$ are matrices, $r$ is a vector. However, it cannot be applied to the non-linear \texttt{SoftMax} in Transformers, which are element-wise operations, not matrix multiplications; hence, a new integrity mechanism is needed. 


To address the above challenges, in this paper, we propose \name{},  to enable a confidential and verifiable Transformer inference. 
The client uploads private data to the cloud server, which performs Transformer inference within the trusted TEEs and untrusted accelerators. \name{}'s protocol enables most Transformer computations to run on accelerators while ensuring data confidentiality and computation integrity. Our protocols and contributions are summarized as follows:
\begin{itemize}
    \item For challenge (I), we design a confidentiality-guaranteed algorithm, \OutAttenMul{}, to securely outsource multiplicative attention operations to an untrusted accelerator. Our algorithm transforms multiplicative linear operations into additive computation with a few pre-computed offline computations, enabling secure outsourcing of these computations.
    
    \item For Challenge (II), we propose a secure \texttt{SoftMax} outsourcing algorithm, \OutSoftMax{}, which offloads its primary computational component (exponentiation) while retaining only a few additions and divisions within TEEs. 
    
 
    \item For Challenge (III), we design \UVerify{} that guarantees the integrity of outsourced computation (both linear and non-linear). For the non-linear \texttt{SoftMax} function particularly, we propose a new \textit{check product} protocol. \UVerify{} also improves efficiency in linear operations compared to prior methods.
    
    \item 
    Through extensive experiments on various models, such as vision, language, and multi-modal Transformers, we show \name{} achieves substantial throughput improvements ranging from  $4.9\times$ to $7.7\times$ for private inferences, and from $3.9\times$ to $6.1\times$ for private verifiable inferences, without sacrificing accuracy. 
\end{itemize}

\section{Threat Model}

We consider an outsourcing scheme between a client-side data owner $\mathcal{C}$ and a server $\mathcal{S}$, where $\mathcal{S}$ executes a Transformer model $f(x): X \rightarrow Y$ on data provided by $\mathcal{C}$. The model $f(\cdot)$ can belong to the server (e.g., in SaaS/API~\cite{achiam2023gpt, medicalGPT, financeGPT}). We adopt a realistic threat model in which the server $\mathcal{S}$ is not fully trustworthy and may be malicious or vulnerable to tampering with the computation results $f(x)$. This departs from the traditional semi-honest setting, in which $\mathcal{S}$ is assumed to be honest but curious about inferring $\mathcal{C}$'s data privacy. An ideal protection scheme should satisfy the following security properties: {\bf Data Privacy}: $\mathcal{S}$ cannot learn any information about input $x$. 
{\bf Integrity \& Verification }: $\mathcal{C}$ could detect an integrity attack when interacting with $\mathcal{S}$ for any input $x$ and ensure the correctness of $y=f(x)$. {\bf Function Privacy}: If $f(\cdot)$ belongs to $\mathcal{S}$, $\mathcal{C}$ cannot learn more about $f(\cdot)$ than what is revealed by $y = f(x)$. Similar to prior works~\cite{tramer2018slalom, hashemi2021darknight}, we assume the availability of TEEs (e.g., Intel SGX) that offer hardware-based data privacy, integrity, and function privacy protection for execution inside an enclave. 
Our methods aims to ensure these security features for computations outside TEEs.
Note that recently Intel SGX has been the subject of side-channel attacks~\cite{van2018foreshadow, van2019ridl, canella2019fallout}, however, most of these issues are being studied with various mitigation techniques proposed~\cite{brasser2019dr, lou2021survey, xu2015controlled}. These attacks are not in the scope of our work. 

\section{Background and Related Work}

\subsection{Transformers}
A basic Transformer consists of an embedding layer and consecutive transformer layers. Every transformer layer is a composition of a multi-head self-attention (MSA) module, a feed-forward (FFN) module, two normalization modules and residual connections. The input data is split into patches, which are then transformed into a token sequence via the embedding layer. The input token sequence can be uniformly denoted as $X_e \in \mathbb{R}^{N \times D}$, where $N$ is the number of tokens and $D$ is the embedding dimension. We describe the main computation blocks in Transformers below.

\noindent \textbf{Additive Linear Operations.} The additive linear operations in Transformers are mainly the linear layers, where the output features are computed by multiplying the input features with weight matrices. For example, in the Attention module, given the input tokens $X_e \in \mathbb{R}^{N\times D}$, the output $Q, K, V \in \mathbb{R}^{N \times d_h}$ are computed by multiplying input $X_e$ with three weight matrices $W_q, W_k, W_v \in \mathbb{R}^{D \times d_h}$, where $d_h$ is the head dimension. Similarly, in the Feed Forward module, the input tokens $X_e \in \mathbb{R}^{N\times D}$ are multiplied by two weight matrices $W_1, W_2 \in \mathbb{R}^{D' \times D}$:
\begin{equation}
\label{e:ffn}
FeedForward(X_e) = Act(X_e\cdot W_1^T + b_1)\cdot W_2 + b_2
\end{equation}
These additive linear operations, such as $(X_e \cdot W_q)$ and $(X_e \cdot W_1^T)$, can be securely outsourced via existing techniques~\cite{tramer2018slalom, hashemi2021darknight}. The additions with the bias matrices $b_1, b_2$ incur only marginal computation overhead in practice. 

\noindent \textbf{Multiplicative Linear Operations.} There are massive multiplicative linear operations in Transformers which cannot be outsourced via prior methods. The primary multiplicative linear operations are computing the attention map and attention output in the Attention module:
\begin{equation}
\label{e:attn}
Attention(Q,K,V) = SoftMax(QK^{T}/\sqrt{d_h})V
\end{equation}
The multiplicative operations (e.g., $Q \cdot K^T$) are fundamentally different from the additive linear operations (e.g., $Q = X_e \cdot W_q$). This is because in the multiplicative operations, neither operand is a constant matrix. As a result, the multiplicative operations cannot be securely outsourced via existing techniques.  We refer to the matrix multiplication between $Q$ and $K$, and between the attention map and $V$ as attention matrix multiplication (AttnMult).

\noindent \textbf{Non-linear Operations.} Apart from the linear operations, Transformers consist of numerous non-linear operations such as the \texttt{SoftMax} function in the Attention module in Equation~\ref{e:attn} and the Activation function \texttt{Act} in Equation~\ref{e:ffn}. These non-linear operations lead to considerable computation overhead during inference. For example, the \texttt{SoftMax} is applied to $(QK^{T}/\sqrt{d_h}) \in \mathbb{R}^{N \times N}$ to compute the attention map. It has a complexity of $O(N^2)$, i.e., quadratic to the input size.

We highlight that the multiplicative linear operations such as $Q \cdot K^T$ and non-linear operations such as \texttt{SoftMax} are computed independently across multiple attention heads. For MSA with $H$ heads, the multi-head attention is computed as:
\begin{equation}
MSA(Q,K,V) = Concat(head_1,...,head_H)W_O
\end{equation}
where $Concat(\cdot)$ is the concatenation operation, 
\begin{equation}
head_i = Attention(XW_q^i,XW_k^i,XW_v^i)
\end{equation}
and $W_O \in \mathbb{R}^{Hd_h \times D}$ is a weight matrix to map features in all heads to the output dimension. The MSA is the key mechanism in the Transformers and also the performance bottleneck. However, existing works cannot securely outsource the heavy computation in the multiplicative linear operations and non-linear operations within the MSA module.

\noindent \textbf{Normalization.} Normalization modules normalize the inputs of MSA and FFN. Given the input tokens $X_e \in \mathbb{R}^{N\times D}$, every value $x_i$ in $X_e$ is normalized to $y_i$ by:
\begin{equation}
y_i = \gamma \cdot ((x_i - \mu) / \sqrt{\sigma^2 + \epsilon}) + \beta
\end{equation}
where $\mu$ is the mean value, $\sigma$ is the standard variance, $\gamma$ is the scaling factor and $\epsilon, \beta$ are offsets. $\mu$ and $\sigma$ are computed differently according to the specific normalization method. Due to their element-wise nature of these operations, it is practical to implement them within TEEs.

\subsection{Trusted Execution Environments (TEEs)}
TEEs like Intel SGX~\cite{sgx} provide a secure environment where data confidentiality and, in some cases, computation integrity are ensured by hardware. Intel SGX specifically safeguards the confidentiality and integrity by isolating data and code within an enclave, shielded from external elements including the operating system, hypervisor, and hardware devices on the system bus. This isolation involves a dedicated memory region, the Processor Reserved Memory (PRM), managed by SGX-enabled CPUs. Here, the Enclave Page Cache (EPC) stores enclave data and code in 4 KB pages, accessible only through specific CPU instructions. This setup prevents unauthorized access to the EPC, maintaining a secure environment for sensitive computations. SGX also supports remote attestation, allowing remote verification of an enclave's integrity through cryptographic proofs. These proofs involve hashing and signing the enclave's contents, verified by Intel's service. This feature has motivated research into running deep learning models within TEEs for security~\cite{hanzlik2021mlcapsule}. Further studies have investigated outsourcing additive linear operations to untrusted accelerators to enhance efficiency without compromising security~\cite{hashemi2021darknight, tramer2018slalom}. Our work expands on this by also outsourcing multiplicative linear operations and non-linear operations, enabling efficient, confidential and verifiable large-scale Transformer inference.

{
\noindent\textbf{Legacy GPUs and GPU-based TEEs.} While some emerging GPUs, such as the NVIDIA Hopper~\cite{choquette2023nvidia}, have begun to support TEE environments, many other emerging GPUs and legacy GPUs (such as GTX series and A100) currently deployed in existing data centers remain in use and are likely to persist for years to come. As a result, CPU-based TEEs are still essential, and our proposed methods can continue to benefit these GPUs effectively. In addition, while GPU TEE can potentially enable native trusted GPU-based model inference, several key issues tamper its realistic adoption. Firstly, only a selected line of GPUs has the feature of TEE, which also needs to be paired with certain CPU TEEs to function properly. Such a configuration is not widely available for real-world deployment. Secondly, many existing large-scale production system are equipped with non-TEE GPUs that still have high performance, upgrading them with the TEE-enabled GPU leads to cost ineffectiveness and sustainability issues. 
Thirdly, with the growing heterogeneity of hardware accelerators (e.g., GPUs, TPUs an FPGAs), a secure computation scheme that relies on each hardware device to support TEE for workload outsourcing is impractical due to the potential compatibility issues among multiple vendors and the complexity of cross-device TEE protocol designs. 
Therefore, designing a secure outsourcing scheme that only utilizes the CPU TEE as the root of trust and can take advantage of the tremendous performance speedup from executions in the untrusted accelerators is imperative. }

\subsection{Secret Sharing for Data Confidentiality}
Secret Sharing~\cite{cramer2015ass, demmler2015ass2, lou2019she} is a cryptographic primitive that allows multiple parties to compute a function over their inputs while keeping them private. All our algorithms are built on a two-party secret sharing over the field $\mathbb{F}_{p}$, where $p$ is a prime number indicating field size. In a two-party secret sharing, a secret $x$ is split into two shares by random sampling $\left \langle x \right \rangle_{0}, \left \langle x \right \rangle_{1} \in \mathbb{F}_{p}$, such that $x =\left \langle x \right \rangle_{0} + \left \langle x \right \rangle_{1} \text{ mod } \mathbb{F}_{p}$. Secret sharing offers a strong security guarantee that, given a share $\left \langle x \right \rangle_{0}$ or $ \left \langle x \right \rangle_{1}$, the value of the original $x$ is hidden, i.e., either party can reconstruct the value of $x$ with negligible possibility~\cite{cramer2015ass}. In the setting of TEE-based confidential inference, the value $x$ can be split by a randomness $r\in\mathbb{F}_{p}$ chosen by the TEEs, such that the two shares are $\left \langle x \right \rangle_{0} = r$ and $\left \langle x \right \rangle_{1}=x-r$, respectively. Prior works~\cite{tramer2018slalom, sun2023shadownet} employ secret sharing to provide privacy guarantees when outsourcing additive linear operation with constant weights \(w\). Yet, existing outsourcing schemes cannot be extended to multiplicative operations where both operands are variables, such as \(Q\) and \(K\), as it is impossible to precompute multiplication between \(r\) and either \(Q\) or \(K\). 


\subsection{Computation Verification for Integrity} 
The verification algorithm enables a client to assert the correctness of computations performed by a server. Within the landscape of TEEs, where computations are outsourced to high-performance untrusted devices such as GPU, ensuring the integrity of these operations is paramount. Soter~\cite{shen2022soter} introduces a "fingerprint" matrix method for integrity checks by the TEEs, which, however, may be vulnerable to targeted attacks. Additionally, recent research~\cite{wei2023v} suggests a sampling-based verification by the TEEs to compare against GPU outputs, facing limitations in detecting selective manipulations without extensive sampling. Freivalds' algorithm~\cite{freivalds1977probabilistic}, referenced in~\cite{tramer2018slalom,sun2023shadownet,hashemi2021darknight}, provides an efficient mechanism for verifying matrix multiplications of the form \(AB=C\). The algorithm commences by generating a random vector \(r\), followed by the TEEs computing the products \(B\cdot r\) and \(C\cdot r\). The next step involves multiplying \(A\) with \(B\cdot r\), and comparing this outcome to \(C\cdot r\). A discrepancy between these products indicates a failure of \(AB\) to equal \(C\), whereas a match suggests a probable equality between \(AB\) and \(C\). Employing this method, the TEEs are able to perform a verification of \(\mathcal{O}(n^3)\) matrix multiplication complexity using a more efficient \(\mathcal{O}(n^2)\) vector-matrix multiplication operation, thereby enhancing the verification efficiency within the TEEs. However, Freivalds' algorithm cannot be used to verify non-linear functions like \texttt{SoftMax}. In contrast, our proposed \UVerify{} method is capable of supporting such verification.

\subsection{Related Work}

In this subsection, we compare \name { } and existing research. The first research direction, denoted by \textit{TEE-only}, focuses on executing all computations within TEEs~\cite{kunkel2019tensorscone, lee2019occlumency}. An example is TensorSCONE~\cite{kunkel2019tensorscone}, which conducts all inference processes inside a TEE enclave to ensure the confidentiality of both the model and data, along with inference integrity. While this approach guarantees security within the enclave, it is less efficient than performing computations in untrusted accelerators outside the TEEs.
The second research trajectory, represented by \textit{additive outsource}, aims to safeguard data confidentiality and inference integrity without necessarily protecting model confidentiality, assuming that the model provider and the cloud server are the same entity. Therefore, there's no need for model confidentiality. This approach, such as Slalom~\cite{tramer2018slalom} and DarKnight~\cite{hashemi2021darknight}, allows for the use of additive confidentiality-preserving and verification techniques to offload certain computations to untrusted hardware. Unlike cloud services that do not require model confidentiality, recent efforts like MLCapsule~\cite{hanzlik2021mlcapsule}, Soter~\cite{shen2022soter}, ShadowNet~\cite{sun2023shadownet} and others~\cite{zhou2023nnsplitter, zhang2023no, liu2023mirrornet, zhang2022teeslice} prioritize model privacy over user input confidentiality in on-device settings, indicating a shift in focus depending on the deployment environment.
The fourth strand of research~\cite{natarajan2023chex, wei2023v} explores enhancing TEEs security through cryptographic methods, such as Fully Homomorphic Encryption (FHE)~\cite{elgamal1985he, paillier1999he2}, to mitigate risks like model theft and side-channel attacks. These enhancements are considered complementary to our approach. Tempo~\cite{xu2024tempo}, which can provide protection for both model and input confidentiality on model training, lacks the established theoretical security foundations seen in Slalom~\cite{tramer2018slalom} and ShadowNet~\cite{sun2023shadownet}. Our work, \name { }, aligns with the second research line but goes beyond their capabilities by facilitating the outsourcing of complex operations like multiplicative attention operations and the non-linear \texttt{SoftMax} function.

\section{Motivation}

We categorize linear operations into two types: \textit{additive} and \textit{multiplicative}. Additive linear operations, involve one \textit{variable} matrix and one \textit{constant} matrix. In contrast, multiplicative linear operations involve operands that are both runtime variables. Prior works show the feasibility of securely outsourcing additive operations to untrusted accelerators via secret sharing and Freivalds' algorithm, which works well for the traditional convolutional neural networks (CNNs) since the convolution of constant pre-trained kernels and variable inputs belongs to the additive operations. We adapt this method to outsource the additive operations in the Transformer, such as $Q=X\cdot W_Q$.
This method involves precomputing $R \cdot W_Q$, outsourcing the operation $(X-R)\cdot W_Q$, and an addition to obtain $Q$ within TEEs. However, this technique does not apply to multiplicative linear operations involving two-variable matrices, for example, $QK^T=(XW_Q) \cdot (XW_K)^T$ in attention multiplication (AttnMult.), which necessitates execution within the TEEs. Additionally, the \texttt{SoftMax} function, essential for creating the attention map from $QK^T$, is inherently non-linear and has not been successfully outsourced by current methods, requiring it to be processed inside the TEEs. Although integrity for linear operations can be confirmed using Freivalds algorithm~\cite{freivalds1977probabilistic}, verifying the integrity of non-linear functions such as \texttt{SoftMax} presents an ongoing challenge.


After relocating the additive linear layers to a GPU, the remaining in-TEE processing of \textit{AttnMult.} and the \texttt{SoftMax} consumes over $60\%$ of the total execution time. This underscores the necessity of externalizing both the multiplicative \textit{AttnMult.} and \texttt{SoftMax} computations to achieve greater efficiency. By outsourcing the multiplicative linear operations, we could potentially halve the execution time required by the TEEs. Moreover, offloading the \texttt{SoftMax} could further reduce latency, with potential savings of up to \(83\%\). While this approach marginally increases the verification workload within the TEEs, the substantial gains in efficiency from outsourcing these operations justify the effort. This evidence motivates the pursuit of secure techniques for offloading attention matrix multiplication and \texttt{SoftMax} computations, with a focus on maintaining computational integrity, particularly for the inherently complex \texttt{SoftMax} operation. 

\section{\name { } Design}
\noindent\textbf{Overview.} 

We first adapt prior work to securely outsource additive linear operations in Transformer. Then we propose \ding{182} \OutAttenMul{} in Section~\ref{sec:method1} to securely outsource multiplicative attention operations. 
With all linear computation outsourced via \OutAttenMul{}, the non-linear \texttt{SoftMax} function becomes the main bottleneck. To this end, we further propose \ding{183} \OutSoftMax{} in Section~\ref{sec:method2}, to outsource the \texttt{SoftMax}. We note that the \texttt{SoftMax} involves a significant number of exponentiations, alongside few additions and divisions. Our strategic approach focuses on outsourcing the expensive exponential computations while retaining the addition and division operations within the TEEs. Specifically, this is achieved by outsourcing \(e^{x+r}\) to accelerators and recovering \(e^x\) by dividing the precomputed \(e^r\). Also, we introduce \ding{184} \UVerify{} in Section~\ref{sec:method3} to ensure the integrity of outsourced computations, especially the verification of non-linear function integrity.

Our experiments reveal that, within the TEEs, normalization and activation functions are relatively lightweight, accounting for less than 5\% of total execution time. Conversely, attention matrix multiplication and \texttt{SoftMax} are identified as primary bottlenecks, consuming approximately 55\% and 35\% of execution time, respectively. By outsourcing these bottlenecks, we can significantly enhance overall efficiency, enabling more efficient and secure Transformer inference.

\begin{figure}[h!]
\centering
\centerline{\includegraphics[width=\linewidth]{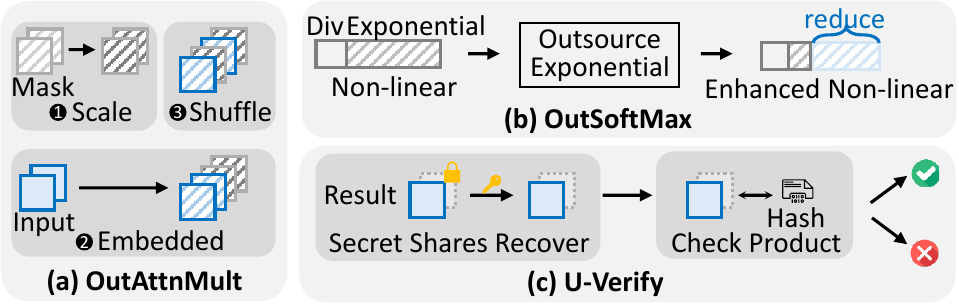}}
\caption{(a) \OutAttenMul{}'s operations within TEEs. (b) \OutSoftMax{} outsources time-consuming exponential calculations in the \texttt{SoftMax} to enhance efficiency. (c) \UVerify{} is performed after recovering the outsourced computation to verify the integrity of the computation by checking the product.}
\label{fig:proposal}
\end{figure}

The starting point is to securely outsource the additive linear operations. We adapt method from prior work~\cite{tramer2018slalom} and integrate our proposed \UVerify{} for more efficient verification. Initially, the input \(X\) is masked with a random matrix \(R\) within the TEEs and sent to accelerators to compute \((X+R)\cdot W\). The TEEs then use the precomputed \(RW\) to recover the desired \(XW\) by subtracting \(RW\) from the accelerators' output.
Verification of outsourced additive operations can be performed using either the traditional Freivalds' algorithm or our proposed \UVerify{}, as detailed in Section~\ref{sec:method3}. By default, we employ \UVerify{}, which, as shown in Figure~\ref{fig:uverify}, provides significant efficiency improvements over the Freivalds' algorithm.

\subsection{Outsource Multiplicative Attention Operation: \OutAttenMul}
\label{sec:method1}
Unlike additive linear operations, multiplicative linear operations involve two variable operands. This variability prevents TEEs from precomputing the product of either operand with the predefined random mask, as they lack prior knowledge about these operands. Specifically, consider the attention multiplication \(Q\cdot K^T\): TEEs (in Secure World) mask \(Q\) with \(R_Q\) and \(K^T\) with \(R_K^T\), then outsource \((Q+R_Q) \cdot (K^T+R_K^T)\) to the accelerators (in Normal World), yielding \(QK^T + R_QK^T + QR_K^T + R_QR_K^T\). To recover the target result \(QK^T\), the TEEs must subtract the additional terms. Among these, only \(R_QR_K^T\) can be precomputed as it does not depend on the variable matrices \(Q\) and \(K^T\), unlike others which cannot be precomputed due to their dependency on \(Q\) or \(K^T\).


We notice that the un-precomputable terms both \(QR_K^T\) and \(R_QK^T\) involve one predetermined mask and one unknown variable operand, allowing their outsourcing via the scheme for additive linear operation. For instance, TEEs can outsource \((Q+R_Q)\cdot R_K^T\) and subsequently obtain \(QR_K^T\) by subtracting the precomputed \(R_QR_K^T\). However, this na\"ive approach presents a critical security risk by exposing \(R^T_K\). This exposure allows the adversary in the normal world to potentially recover \(K^T\) using \(K^T+R_K^T\) obtained from the initial outsourcing round. 

To prevent this risk, we propose a strategy that enhances security by using a scaled version \(bR_K^T\) rather than \(R_K^T\). More importantly, \(bR_K^T\) is not transmitted directly to the accelerator but is integrated into the matrix \(K^T+R_K^T\) through a column-wise permutation. This approach serves two primary security functions: 1) it conceals the distinction between \(K^T+R_K^T\) and \(bR_K^T\), thwarting attackers from identifying them, and 2) it facilitates the simultaneous computation of \((Q+R_Q)\cdot bR_K^T\) along with \((Q+R_Q)\cdot (K^T+R_K^T)\), thereby obviating the need for an additional round of outsourcing. Subsequently, the TEEs can retrieve \(QR_K^T\) from \((Q+R_Q)\cdot R_K^T\) by applying scalar multiplication with \(1/b\) and subtracting \(R_QR_K^T\). Although we focus here on \(K^T\), the processing of \(Q\) employs a similar principle.

\begin{figure}[t!]
\centering
\centerline{\includegraphics[width=\linewidth]{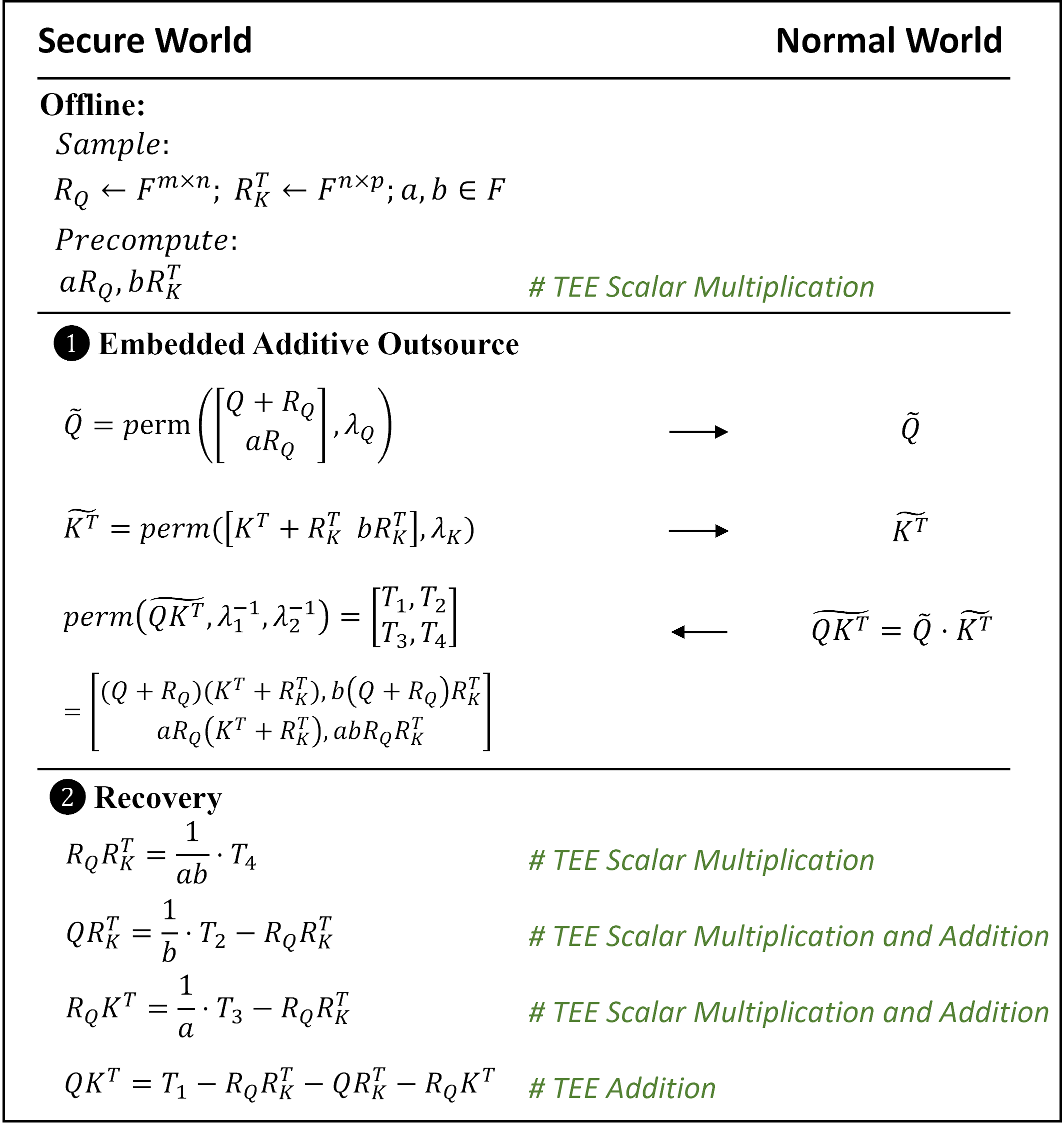}}
\caption{Illustration of Outsource Multiplicative attention operation: \OutAttenMul{}}
\label{fig:method1}
\end{figure}

The details of \OutAttenMul{} are in Figure~\ref{fig:method1}. Given input matrices \( Q\in \mathbb{F}^{m\times n} \) and \( K^T\in \mathbb{F}^{n \times p} \) in a finite filed \(\mathbb{F}\), \OutAttenMul{} is divided into offline phase and online phase.

\noindent\textbf{Offline Preprocessing.} Initially, TEEs (in Secure World) generate two random matrices \( R_Q\in \mathbb{F}^{m \times n} \) and \( R_K^T \in \mathbb{F}^{n \times p} \). It then precomputes \(aR_Q\) and \(bR_K^T\) by two scalar multiplications.

\noindent\textbf{Embedded Additive Outsource.} In this stage, TEEs first obfuscate \(Q\) and \(K^T\) to \(Q+R_Q\) and \(K^T+R_K^T\), respectively. These matrices are then embedded into \(\widetilde{Q}\) and \(\widetilde{K^T}\) through strategic permutations. Specifically, \(\widetilde{Q}\) is crafted by vertically stacking \(aR_Q\) beneath \(Q+R_Q\) and applying a row-wise permutation,
\begin{equation}
\footnotesize
    \widetilde{Q} = perm\left(\begin{bmatrix}
    Q + R_Q \\
    aR_Q 
    \end{bmatrix}, \lambda_1\right)
\label{e:Q_construction}
\end{equation}
Similarly, \(\widetilde{K^T}\) is constructed by horizontally concatenating \(K^T+R_K^T\) with \(bR_K^T\) and applying a column-wise permutation,
\begin{equation}
    \widetilde{K^T} = perm\left(\begin{bmatrix}
    K^T + R_K^T & bR_K^T
    \end{bmatrix}, \lambda_2\right)
\label{e:K_construction}
\end{equation}
\( perm(\cdot, \lambda) \) here indicates matrix permutation with permutation indices \( \lambda \), so that the adversary in normal world cannot distinguish \(R_Q\) or \(R_K^T\) from the blinded matrics.

These blinded matrices are then outsourced to the accelerator (in Normal World) for multiplication. After recovering the received results with the permutation indices, TEEs get:

\begin{equation}
\label{e:gpu_mul}
\small
perm(\widetilde{QK^T}, \lambda_1^{-1}, \lambda_2^{-1}) = 
\begin{bmatrix}
(Q + R_Q)(K^T+R_K^T) & a(Q + R_Q)R_K^T \\
bR_Q(K^T+R_K^T) & abR_QR^T_K
\end{bmatrix}
\end{equation}
Figure~\ref{fig:proposal} (a) intuitively shows this masked input processing.

\noindent\textbf{Recover.} As detailed in Figure~\ref{fig:method1}, the TEEs start with applying a scalar multiplication to \(abR_QR_K^T\) to obtain \(R_QR_K^T\). The TEEs then retrieve \(QR_K^T\) and \(R_QK^T\) by performing scalar multiplications on \(a(Q + R_Q)R_K^T\) and \(bR_Q(K^T+R_K^T)\), respectively, and subsequently subtracting \(R_QR_K^T\) from each. The final recovery of \(QK^T\) is achieved by strategically subtracting these terms.

\noindent\textbf{Complexity Analysis.} In vanilla secure matrix multiplication within TEEs, computing \(QK^T\) for matrices \(Q\in \mathbb{F}^{m\times n}\) and \(K^T \in \mathbb{F}^{n\times p}\) requires \(\mathcal{O}(mnp)\) multiplications to be performed in resource-constrained TEEs. In contrast, \OutAttenMul{} significantly reduces this burden by offloading the bulk of computation to the accelerators.

In the offline phase, TEEs perform two scalar multiplications with a complexity of \(O(mn+np)\) for \(aR_Q\) and \(bR_K\).

At the Embedded Additive Outsource stage, TEEs execute two permutations and two additions to prepare \(\widetilde{Q}\) and \(\widetilde{K^T}\). The GPU then handles the computationally intensive matrix multiplication \(\widetilde{Q}\cdot \widetilde{K^T}\), with a complexity of \(\mathcal{O}(mnp)\), given that \( \widetilde{Q} \in \mathbb{F}^{2m \times n}\) and \( \widetilde{K^T} \in \mathbb{F}^{n \times 2p}\).

Finally, in the recovery stage, the TEEs perform three scalar multiplications with \(O(mn+np)\) and five additions to recover the desired \( QK^T \). Overall, \OutAttenMul{} shifts the computational load from \( \mathcal{O}(mnp) \) multiplications within the TEEs to \( \mathcal{O}(mnp) \) multiplications on the accelerator, alongside scalar multiplications and less costly permutation, and addition operations within the TEEs.

\noindent\textbf{Security Analysis.} 
In the outsourcing protocol in Figure~\ref{fig:method1}, data within the TEEs (the Secure World) is protected, while data processed in accelerators (the Normal World) is exposed to potential attackers. Our goal is to prevent the attackers in the normal world from deducing the original \(Q\) or \(K^T\). To achieve this, the TEEs construct \(\widetilde{Q}\) and \(\widetilde{K^T}\) via Equations~\ref{e:Q_construction} and \ref{e:K_construction}. Taking \(\widetilde{Q}\) as an example, the TEEs create a secret share by adding \(R_Q\) to \(Q\) and subsequently permutes \(Q+R_Q\) together with \(aR_Q\) using private permutation indices \(\lambda_1\). Since \(Q+R_Q\) is equivalent to applying a one-time pad~\cite{bellare2001introduction}, its distribution is indistinguishable from \(aR_Q\) in the view of attackers~\cite{zhang2023no}.
The security level is quantified as $\log (d\cdot (2m)!)$, where \(2m\) represents the total rows in \(\widetilde{Q}\) and \(d\) denotes the finite field size. This security level estimates the probabilities for an attacker to accurately discern \(Q+R_Q\) and \(aR_Q\) from \(\widetilde{Q}\) and to correctly identify scalar \(a\). In Transformers, \(Q\) typically has a large dimension (e.g., 128 for BERT), an 8-bit scalar would provide a security level of approximately 13,471 bits. Additionally, by expanding \(R_Q\) with random values and assigning varying scalars to different rows (columns for \(R_K^T\)), the TEEs can tailor the security level to meet specific requirements and matrix sizes, further enhancing protection. Detailed methodologies and additional insights are presented in Appendix~\ref{app:security_attenmul}.

\subsection{Outsource Non-linear \texttt{SoftMax}: \\ \OutSoftMax{}}
\label{sec:method2}

Prior outsourcing methods for CNN-based models typically offload linear layers to accelerators, while keeping non-linear \texttt{ReLU} within TEEs due to their relatively simpler computations and the difficulties of non-linear outsourcing. However, in the context of Transformer models, the \texttt{SoftMax} within attention layers poses a substantial computational bottleneck, accounting for about 64\% of the total processing time after linear operations have been outsourced from TEEs for sequences of length 512. The complexity of the \texttt{SoftMax} operation increases quadratically with the input length, which means its computational burden becomes even more pronounced. Specifically, within the \texttt{SoftMax} process, the exponentiation operation alone is responsible for 92.9\% of the \texttt{SoftMax} inference time when executed within the TEEs.


The rationale for outsourcing the \texttt{SoftMax} function stems from its reliance on extensive exponential calculations, which interestingly exhibit a property akin to linear operations. Specifically, linear layers are amenable to outsourcing due to the distributive property of matrix addition over multiplication, facilitating the computation of $(X + R) \cdot W$ as $XW + RW$. In contrast, non-linear layers typically do not share this trait. Yet, the exponential function crucial to the \texttt{SoftMax} displays a similar linear-like property: $e^{X + R} = e^X \cdot e^R$. By precalculating $e^R$ during the offline phase, the TEEs can securely outsource the exponential computation $e^{X+R}$, capitalizing on this linear-like behavior. The accelerator then processes $e^{X+R}$ and sends it back to the TEEs, where $e^X$ is derived by dividing $e^{X+R}$ by the precalculated $e^R$ on an element-wise basis. This method effectively transforms an exponential operation into a multiplication, substantially easing the computational load as shown in Figure~\ref{fig:proposal} (b). This is especially advantageous given that exponentiation is significantly more resource-intensive compared to basic arithmetic operations within the TEEs.

\begin{figure}[ht!]
    \centering
    \centerline{\includegraphics[width=0.9\linewidth]{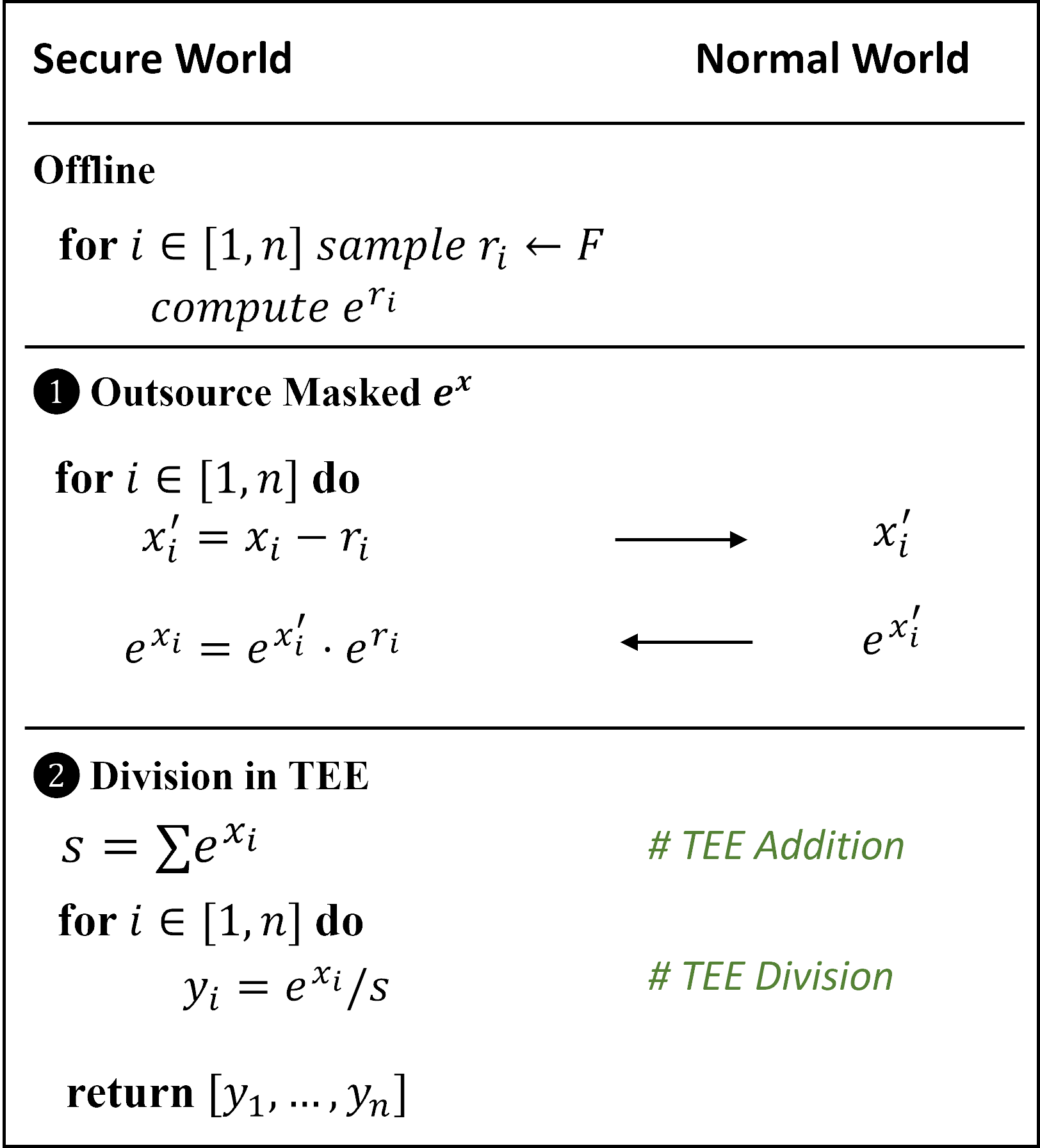}}
    \caption{Outsourcing non-linear \texttt{SoftMax}: \OutSoftMax{}}
    \label{fig:method2}
\end{figure}

Our \OutSoftMax{} algorithm is depicted in Figure~\ref{fig:method2} and comprises one offline stage along with two online stages: Outsource Masked \(e^X\) and Division in the TEEs. For an input vector \(X \in \mathbb{F}^n\), the process is as described below.

\noindent\textbf{Offline Preprocessing.} During the offline stage, for each element \(x_i\) of the input \(X\), the TEEs in the secure world sample a corresponding random value \(r_i\) from the field \(\mathbb{F}\) and computes:

\begin{equation}
    e^{r_i}, \quad \forall i \in \{1, \dots, n\}
\end{equation}

\noindent\textbf{Outsource Masked \(e^X\).} TEEs mask the input vector by computing:

\begin{equation}
x'_i = x_i - r_i, \quad \forall i \in \{1, \dots, n\}
\end{equation}
Then send \(x'_i\) to the accelerator in normal world, which returns:

\begin{equation}
e^{x'_i}, \quad \forall i \in \{1, \dots, n\}
\end{equation}
Upon receving the results, the TEEs restore \(e^{x_i}\) by multiplying the accelerator's output with the precomputed exponentials:

\begin{equation}
e^{x_i} = e^{x'_i} \cdot e^{r_i}, \quad \forall i \in \{1, \dots, n\}
\label{e:recover_softmax}
\end{equation}

\noindent\textbf{Division in the TEEs. } TEEs firstly compute the normalization scalar by:

\begin{equation}
s = \sum_{i=1}^n e^{x_i}
\end{equation}
and subsequently, the \texttt{SoftMax} scores are computed by divisions:

\begin{equation}
y_i = \frac{e^{x_i}}{s}, \quad \forall i \in \{1, \dots, n\}
\end{equation}
which completes the secure \texttt{SoftMax} outsourcing.

By leveraging the outsourced computation for the most resource-demanding operation, i.e., exponentiation, the \OutSoftMax{} efficiently computes \texttt{SoftMax} scores within the TEE's constraints.


\noindent\textbf{Complexity Analysis.} In the offline phase, Outsource Masked \(e^X\), the TEEs prepare for the \texttt{SoftMax} operation by sampling and precomputing the exponentiations of \( n \) random values, one for each element of the input vector.
During the first online phase, the TEEs mask each input element by performing \( n \) subtractions. The masked values are then outsourced to the accelerator for further processing, which undertakes \( n \) exponentiation operations.
In the online phase, i.e., Division in the TEEs, the TEEs complete the \texttt{SoftMax} computation by executing \( n \) multiplications to combine the accelerator's output \(e^{x_i'}\) with precomputed values \(e^{r_i}\), followed by \( n \) additions to sum the exponential terms and \( n \) multiplications to calculate the \texttt{SoftMax} probabilities. So our \OutSoftMax{} convert \(n\) exponentiation operations to convert \(2n\) multiplication operations and \(n\) addition operations in the TEEs.
By relocating the exponentiation tasks to the accelerators, the \OutSoftMax{} algorithm relieves the TEEs of the most computationally demanding aspect of the \texttt{SoftMax} function. Considering that exponentiation can be an order of magnitude more time-consuming than simpler arithmetic operations in the TEEs setting, outsourcing these \( n \) operations significantly enhances the efficiency of secure inference processes. The strategic offloading of these tasks ensures that the performance bottleneck within the TEEs is mitigated.

\noindent \textbf{Security Analysis.} In our \OutSoftMax{} protocol, only the transformed vector \(X'=[x_1-r_1, ..., x_n-r_n]\) is exposed to the normal world. Due to the additive secret sharing~\cite{cramer2015ass, demmler2015ass2}, \(X'=X-R\) can be viewed as one of the secret shares of original \(X\). Without the other share \(R=[r_1, ..., r_n]\), the attacker cannot reconstruct the original \(X\) from \(X'\).

\subsection{\UVerify{}}
\label{sec:method3}

\noindent\textbf{Verification of \OutSoftMax{}.} Prior matrix multiplication outsourcing schemes use Freivalds' algorithm~\cite{freivalds1977probabilistic} to verify whether the the product of two input matrices \(A\cdot B\) equals to the output matrix \(C\). It achieves this by multiplying both \(B\) and \(C\) by a random vector \(r\), then checking if \(A\) times the result of \(Br\) equals \(Cr\). In this case, the TEEs can use \(\mathcal{O}(mn+np)\) multiplications to verify an \(\mathcal{O}(mnp)\) matrix multiplication, where \((m,n)\) is the size of matrix $A$ and \((n,p)\) is the size of matrix $B$. However, Freivalds' algorithm is unsuitable for verifying element-wise operations like exponentiation since it relies on matrix-specific properties. In the \texttt{SoftMax} computation, exponentiation is applied individually to each element, lacking the associative and distributive properties that Freivalds' algorithm depends on, thus necessitating a different approach for verification. Our insight stems from recognizing the unique \textit{linear-like feature} of the exponential function where \(e^{a_1x_1+a_2x_2}=(e^{x_1})^{a_1}\cdot (e^{x_2})^{a_2}\).

\begin{figure}[ht!]
    \centering
    \centerline{\includegraphics[width=0.85\linewidth]{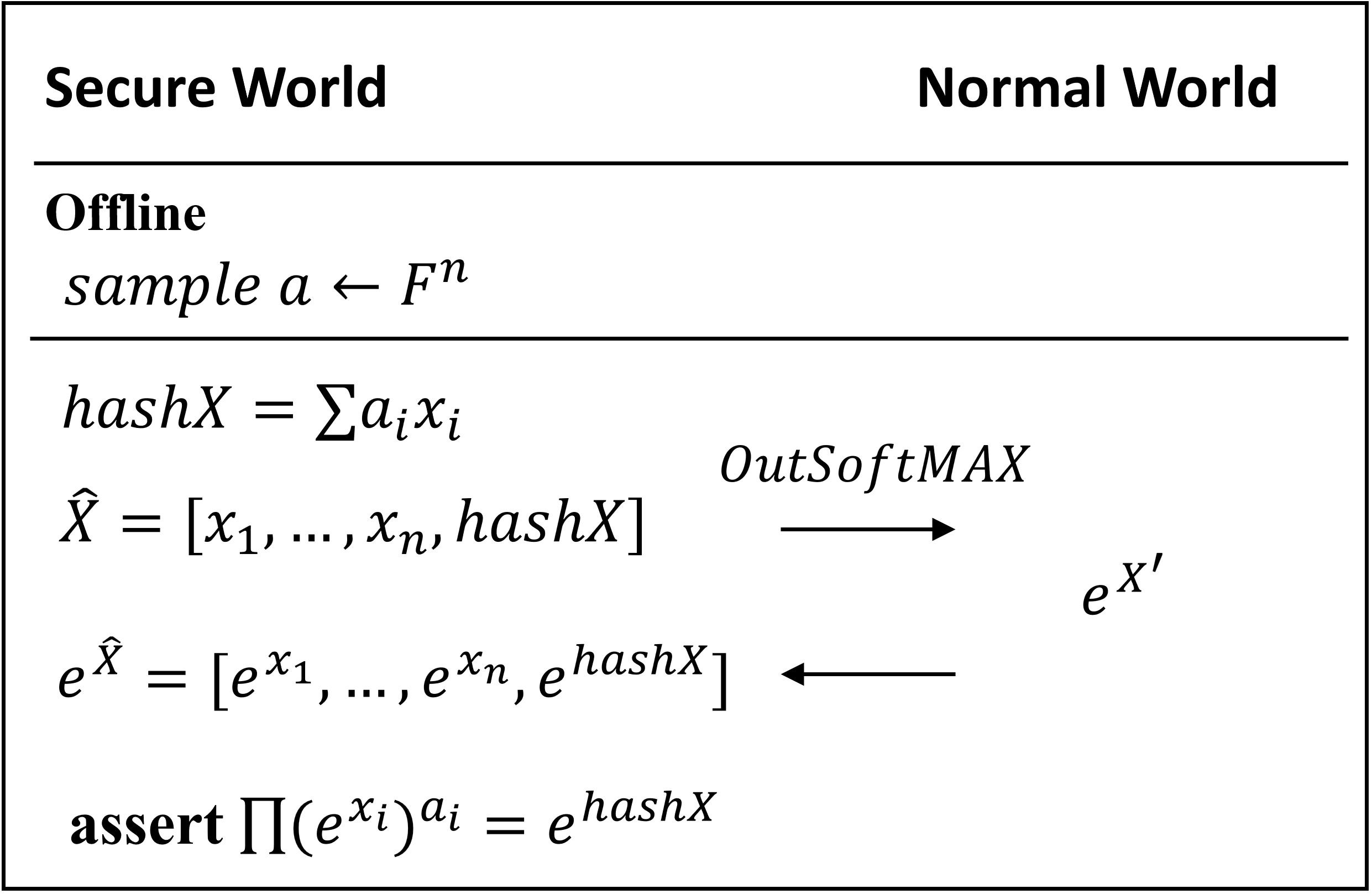}}
    \caption{\UVerify{} on \OutSoftMax{}.}
    \label{fig:method3a}
\end{figure}

Figure~\ref{fig:method3a} illustrates the procedure of our verification method for \OutSoftMax{}. Consider an input vector \( X = [x_1, ..., x_n] \). The TEEs begin with generating a secret random vector \(a = [a_1, ..., a_n]\), and computing the hash of \(X\), denoted as \( hashX \), using the equation: $hashX = \sum_{i=1}^{n} a_i x_i$. This hash is then randomly inserted into \( X \) to form \(\hat{X}\), and then, TEEs utilize \OutSoftMax{} to send accelerators the secret share, $X'=\hat{X}-R= [x_1-r_1,...,hashX-r_{hash},...,x_n-r_n] = [x_1',...,hashX',...,x_n']$ to compute the exponential of each element, resulting in: $e^{X'} = [e^{x_1'}, ..., e^{hashX'},..., e^{x_n'}]$. Upon receiving \( e^{X'} \) from accelerators, the TEEs first recover \(e^X\) according to Equation~\ref{e:recover_softmax} and then verifies the correctness of the computation by checking the following equality: 

\begin{equation}
\prod_{i=1}^{n} (e^{x_i})^{a_i} = e^{hashX}
\label{e:hash_softmax}
\end{equation}

As depicted in Figure~\ref{fig:proposal} (c), since the hash is integrated prior to the outsourcing mask, integrity verification is performed after the results are recovered.

\noindent\textbf{Complexity Analysis.} In the \texttt{SoftMax} verification process, the TEEs first generate random positive coefficients \(a = [a_1, ..., a_n]\) and computes the \(hashX = \sum a_i x_i\), involving \(n\) multiplications and \(n-1\) additions. This hash, alongside \(X\), forms an augmented vector \(\hat{X}\), which is then outsourced by \OutSoftMax{} to the accelerator to compute \(e^{X'}\). The TEEs subsequently verify the computation by comparing \(\prod (e^{x_i})^{a_i}\) with \(e^{hashX}\), ensuring the correctness of the \texttt{SoftMax} operation. This verification requires up to \(n(a_{\text{max}} - 1) + n\) multiplications, depending on the largest \(a_i\) value. If the TEEs constrain the values of \(a\) below $3$, the required multiplications are capped at \(3n\), making this verification process significantly more efficient than performing \(n\) exponentiations directly within the TEEs for \texttt{SoftMax} computation.

\noindent\textbf{Security Analysis.}
For a successful adversarial scenario, where the attackers seek to alter the \(e^{x_i}\) element in the correct output \(e^{X}\), they would need to modify the $hashX$ according to the coefficient \(a_i\) to bypass the verification. However, without knowledge of both coefficients \(a_i\) and the location of \(hashX\) within \(\hat{X}\), the attack success rate is limited to \(\frac{1}{n \cdot 2^d}\), where \(n\) is the size of \(e^X\), and \(2^d\) represents the space of possible \(a_i\). Formally, the security level is expressed as $\log (n\cdot 2^d)$. 

More importantly, since \OutSoftMax{} is applied to the hashed vector \(\hat{X}\), which is blinded with a random mask, the security of \UVerify{} is further strengthened by \OutSoftMax{}. In other words, to launch an attack, the attackers must first brute-force the random masks used in \OutSoftMax{} to recover \(\hat{X}\) from \(X'\). Thus, the overall security level is raised to $\log (n\cdot 2^d\cdot d^n)$, where $\frac{1}{d^n}$ represents the probability that attackers correctly identify all masks used in \OutSoftMax{}. 

Furthermore, traditional replay attacks are thwarted by the one-time use of scalars in \UVerify{} and masks in \OutSoftMax{}. More details can be found in Appendix~\ref{app:security_uverify_softmax}.

\begin{figure}[ht!]
    \centering
    \centerline{\includegraphics[width=0.85\linewidth]{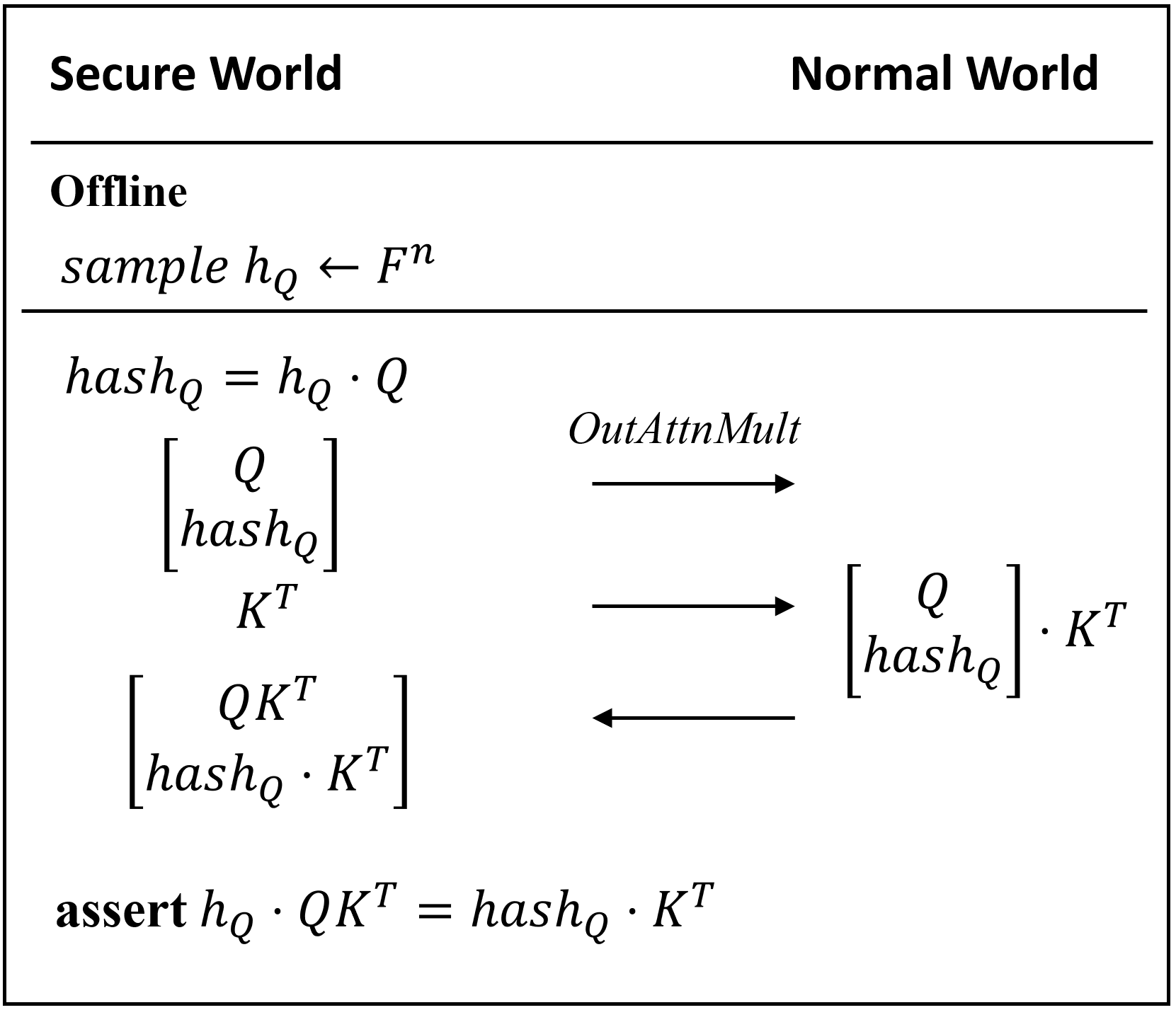}}
    \caption{\UVerify{} on \OutAttenMul{}.}
    \label{fig:method3b}
\end{figure}

\noindent\textbf{Verification of \OutAttenMul{}.} Figure~\ref{fig:method3b} outlines the procedure for verifying \OutAttenMul{} operations. In the offline phase, the TEEs commence by sampling a secret random vector \( h_Q \) from the finite field \( \mathbb{F}^n \) to compute the hash \( hashQ \) as follows: $hashQ = h_Q \cdot Q $. Next, during the online phase, the TEEs construct an augmented matrix by appending \( hashQ \) to matrix \( Q \), and then utilizes the accelerator's computation of the product with \( K^T \) using \OutAttenMul{}. Upon receiving the results from the accelerator: 
\begin{equation}
    \begin{bmatrix} Q \\ hashQ\end{bmatrix} \cdot K^T =\begin{bmatrix} QK^T \\ hashQ\cdot K^T\end{bmatrix}
\label{e:verification_attention}
\end{equation}

The TEEs recover the result via \OutAttenMul{}, and then verify the multiplication's correctness by ensuring that \( h_Q \cdot QK^T \) equals \( hashQ \cdot K^T \).

\noindent\textbf{Security Analysis.} 
A successful integrity attack requires tampering with the matrix multiplication result \(QK^T\) such that it still passes the integrity check \(h_Q \cdot QK^T = hash_Q \cdot K^T\). To bypass this check, an attacker would need to modify \(hash_Q\) so that the altered \(hash_Q \cdot K^T\) remains consistent with \(h_Q \cdot QK^T\). However, multiple barriers prevent such an attack. First, the permutation procedure in \OutAttenMul{} obscures \(hash_Q\) within a mixture of \(hash_Q\) and \(Q\), making it impossible for the attacker to identify \(hash_Q\). Second, the additive masks applied during \OutAttenMul{} render any malicious multiplication ineffective. Most critically, the random vector \(h_Q\) is available only within the TEEs, so without knowing \(h_Q\), it is impossible to maintain the equality \(h_Q \cdot QK^T = hash_Q \cdot K^T\) by modifying \(QK^T\), \(hash_Q\), or \(K^T\). Simply multiplying all terms by the same scalar also fails because the mask introduced by \OutAttenMul{} disrupts this multiplicative relationship.
More discussions can be found in Appendix~\ref{app:security_uverify_atten}.

\noindent\textbf{Complexity Analysis.} The TEEs compute the hash \( hashA \) using \( n \times m \) multiplications and few additions. The accelerator then performs the bulk of computation by multiplying the augmented matrix with \( B \), requiring a total of \( (m+1) \times n \times p \) multiplications for both the matrix product and the hash. Upon receiving the result, the TEEs verify the integrity with a single vector-matrix multiplication involving \( m \times p \) multiplications.

\section{Experimental Methodology}
\label{sec:setup}
\noindent In this section, we introduce the experimental methodology.

\noindent\textbf{Models.} In our evaluation, we selected four transformers: 1) ViT-16B~\cite{dosovitskiy2020image} applies the transformer structure to computer vision, comprised of $12$ layers and 16 attention heads, each with a hidden feature size of $768$, tailored for image classification. 2) BERT-Base~\cite{devlin2018bert}, a seminal model in NLP, features $12$ layers and $12$ attention heads, each with a hidden feature size of $768$, aimed at comprehending and processing language. 3) LLaMA-7B~\cite{touvron2023llama} marks an advancement in language models with its $32$ layers and $32$ attention heads, each with an extensive hidden feature size of $4096$, for intricate language tasks. 4) CLIP~\cite{radford2021learning} fuses vision and language by employing $12$ layers with $12$ attention heads for both its text and visual encoders, each with a hidden feature size of $768$, and is trained across a diverse set of images and text pairs.

\noindent\textbf{Datasets.} ViT-16B and CLIP are evaluated on the ImageNet~\cite{deng2009imagenet} dataset, a benchmark for image classification, with accuracy serving as the metric for success. BERT-Base is tested against the SST-2 dataset~\cite{socher2013recursive}, a standard for sentiment analysis in NLP, also using accuracy as the evaluative measure. For LLaMA-7B, we employ the Wiki-Text dataset~\cite{merity2016pointer} to gauge its language modeling capability, utilizing Perplexity as the metric, which quantifies how well a probability model predicts a sample. Accuracy measures the proportion of correct predictions over the total, reflecting classification performance, while Perplexity measures the model's certainty in its predictions, with lower values indicating better predictive performance.

\noindent\textbf{System Setup and Implementation.} We conducted the \name { } implementation on a server powered by an Intel(R) Xeon(R) Gold 6342 CPU, operating at 2.8GHz, and equipped with 512GB of DRAM. This setup also included an NVIDIA A40 GPU with 48GB of VRAM. Our SGX implementation leveraged Eigen~\cite{eigenweb}, a linear-algebra library also employed by TensorFlow for constructing DNN layers such as the attention module, \texttt{SoftMax}, LayerNorm, GeLU, and ReLU. The development environment included TensorFlow and Python 3, used for model quantization and inference. We sourced pre-trained models for ViT, BERT, and CLIP from Keras~\cite{chollet2015keras} and applied quantization techniques as described in recent studies~\cite{liu2021post, kim2021bert}. Additionally, we acquired a pre-trained, quantized 8-bit LLaMA model from HuggingFace~\cite{hugging_face}.



\noindent\textbf{Quantization.} 
\name { } adopts a quantization strategy for both inputs and model weights, drawing on the approaches of Slalom~\cite{tramer2018slalom} and DarKnight~\cite{hashemi2021darknight}. Initially, it converts values from floating-point to fixed-point by selecting a fractional bit number, \(l\), scaling values by \(2^l\), and rounding to integers. For negative values, a correction \(p\) is applied to adjust them into the field \(\mathbb{Z}_p\), where prime \(p=2^{24}-3\). Subsequent computations are outsourced to the GPUs by the TEEs, which later de-quantizes the GPU's results to obtain the original values. Our experimental setup, with \(l=8\), resulted in a maximum accuracy drop of 1.9\% as shown in Table~\ref{t:quantize}. Accuracy for ViT and CLIP was measured on the ImageNet validation set, BERT-Base on the SST-2 dataset, and LLaMA's perplexity on the Wiki-Text dataset, experiencing at most a 1.9\% accuracy decrease and a 0.21 increase in perplexity for LLaMA.


\section{Experimental Results}
\subsection{End-to-end performance}

\noindent\textbf{Comparison to baseline methods.} In Table~\ref{t:end2end}, we provide a detailed evaluation of \name { } across four transformer architectures and three distinct datasets, both with and without integrity verification. We first compare against a TEE-only baseline~\cite{hanzlik2021mlcapsule}, where all inferences are executed entirely within the TEEs, demonstrating the least efficiency. For example, a single ViT inference takes \(0.713\) s in this setup. By outsourcing additive matrix multiplications between input \(X\) and weights \(W\), an approach labeled as "Additive OutSrc.," we achieve a \(1.5\times\) average speedup with integrity verification. 

Building on this, \name { } significantly extends the outsourcing capabilities by offloading all multiplicative linear operations and \texttt{SoftMax} computations in the attention module from TEEs to the GPU. This optimization results in a \(3.6\times\) speedup over "Additive OutSrc." and an overall \(5.4\times\) improvement compared to the TEE-only baseline, while maintaining both integrity and privacy. These enhancements are primarily due to \name { }’s efficient outsourcing of computational bottlenecks, particularly multiplicative matrix multiplications and \texttt{SoftMax}. With only privacy protection (and no matrix multiplication required for verification), \name { } achieves an even greater \(6.67\times\) average speedup.

The performance gains are especially pronounced in larger models like LLaMA, where \name { } achieves a \(6.1\times\) speedup over the TEE-only baseline. This highlights the critical role of GPU-based outsourcing in mitigating the resource limitations of TEEs, particularly for expansive transformer models.

\begin{table}[ht!]
\centering
\setlength{\tabcolsep}{2pt}
\caption{\textcolor{black}{Comparison of \name { } and prior methods, i.e., TEE-only and additive linear outsourcing methods, on end-to-end time (s).}}
\begin{tabular}{llcccc}\toprule
\multirow{2}{*}{Method} & & \multirow{2}{*}{\makecell[c]{ViT-\\ImageNet}} & \multirow{2}{*}{\makecell[c]{BERT-\\SST2}} & \multirow{2}{*}{\makecell[c]{CLIP-\\ImageNet}} & \multirow{2}{*}{\makecell[c]{LLaMA-\\WiKi}} \\
& & & & \\
\midrule\midrule
\multicolumn{2}{l}{TEE-only} & 0.713 & 1.294 & 1.972 & 113.4 \\\midrule
\multirow{2}{*}{Additive OutSrc.} & w/o verf. & 0.511 & 0.858 & 1.255 & 72.53  \\
 & w/ verf. & 0.523 & 0.886 & 1.298 & 75.28  \\\midrule
\multirow{2}{*}{\name { }} & w/o verf. & 0.151 & 0.174 & 0.205 & 14.77 \\
 & w/ verf. & 0.188 & 0.216 & 0.363 & 18.60 \\\bottomrule
\end{tabular}
\label{t:end2end}
\end{table}

\begin{figure}[h!]
    \centering
    \includegraphics[width=\linewidth]{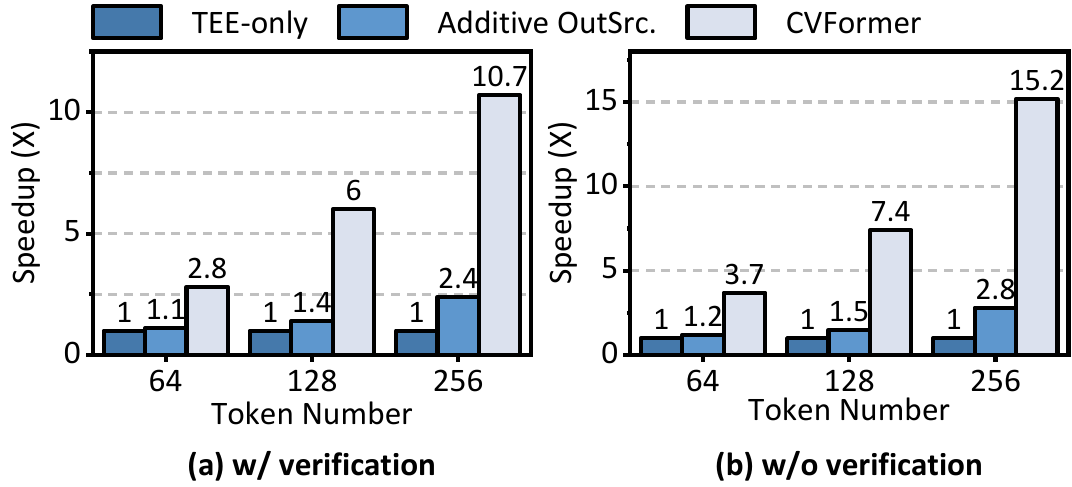}
    \caption{\name { } obtains higher performance speedup on long-token inputs.}
    \label{fig:token_ablation}
\end{figure}

\noindent\textbf{Results on long-token inputs.} Table~\ref{t:end2end} illustrates that \name { } achieves a \(6.0\times\) speedup on BERT with an input token number of \(128\). The performance improvement can be further pronounced when the inputs have more tokens (long-token inputs). The Figure~\ref{fig:token_ablation} (a) and (b) depict the speedup with and without integrity verification, respectively. The results showcase that \name { }'s advantage over TEE-only and "Additive OutSrc." methods grows with increasing token numbers. For instance, with integrity verification, the speedup with \(64\) input tokens is \(2.8\times\) and surges to \(10.7\times\) with \(256\) tokens. This is attributed to \name { }'s optimization strategy, which simplifies the TEEs' workload from \(\mathcal{O}(n^3)\) to \(\mathcal{O}(n^2)\) matrix multiplications and transforms \(\mathcal{O}(n)\) exponential operations to multiplications. Without the integrity verification, the speedup is even more pronounced, reaching \(15.2\times\) for \(256\) tokens, as it eliminates the need for the TEEs to perform \(\mathcal{O}(n^2)\) matrix multiplications for verification. Conversely, the "Additive OutSrc". approach does not achieve such significant gains because it still relies on TEEs for the remaining matrix multiplication and \texttt{SoftMax} computations, which become bottlenecks and limit performance improvements across varying token numbers.


\noindent \textbf{Results on standard CPU.} To assess our outsourcing schemes' performance without the specific constraints of Intel SGX, we evaluated the benchmarks presented in Table~\ref{t:end2end} on the same CPU, but outside SGX enclave mode. Table~\ref{t:cpu} demonstrates benchmarks for BERT and ViT on a single core with either direct computation or various secure outsourcing schemes. On the BERT, by outsourcing additive matrix multiplication to the GPU (Additive OutSrc.) with integrity verification, the latency drops from \(0.492\) s to \(0.334\) s. This is further reduced to \(0.131\) s with \name { }, which outsources both linear and non-linear operations, achieving a \(3.8\times\) speedup. For ViT, the \name { } can achieve \(3.3\times\) and \(4.1\times\) speedup with and without integrity verification.
\begin{table}[ht!]
\centering
\setlength{\tabcolsep}{9pt}
\caption{\textcolor{black}{Comparison of \name { } and prior methods using BERT and ViT on an untrusted CPU. }}
\begin{tabular}{llcc}\toprule
\multicolumn{2}{l}{Method} & BERT (s) & ViT (s) \\\midrule\midrule
\multicolumn{2}{l}{CPU-only} & 0.492 & 0.331 \\\midrule
\multirow{2}{*}{Additive OutSrc.} & w/o verf. & 0.318 & 0.235 \\
 & w/ verf. & 0.334 & 0.247 \\\midrule
\multirow{2}{*}{\name { }} & w/o verf. & 0.095 & 0.081 \\
 & w/ verf. & 0.131 & 0.102 \\\bottomrule
 
\end{tabular}
\label{t:cpu}
\end{table}

\subsection{Ablation Study and Benchmark}
\noindent\textbf{Ablation study on the effectiveness of proposed techniques.} Table~\ref{t:ablation} evaluates the performance of our proposed methods on the BERT model with a 128-token input, comparing scenarios both with and without integrity verification. With integrity verification enabled, outsourcing additive matrix multiplications reduces the inference latency from \(1.294\) s to \(0.886\) s. Further optimization is achieved by utilizing \OutAttenMul{} to outsource multiplicative attention matrix multiplications, which significantly decreases the end-to-end latency to \(0.570\) s. This improvement is attributed to simplifying the TEEs' workload from matrix-matrix to vector-matrix multiplications for validating GPU computations. 

In addition, incorporating \OutSoftMax{} to outsource the \texttt{SoftMax} operation further reduces the latency to \(0.531\) s. This reduction reflects the shift in the TEE's role from performing computationally intensive exponential operations to simpler multiplication tasks. By combining both outsourcing strategies, the inference latency is ultimately reduced to just \(0.216\) s, achieving a substantial \(83.3\%\) reduction for a single BERT inference.

\begin{table}[ht!]
\centering
\setlength{\tabcolsep}{2pt}
\caption{Ablation study of proposed techniques. TEE-only natively provides integrity guarantee.}
\begin{tabular}{lcc}\toprule
\multirow{2}{*}{Technique} & \multicolumn{2}{c}{Time (s)} \\\cmidrule(lr){2-3}
& w/o verf. & w/ verf. \\\midrule\midrule
TEE-only & - & 1.294 \\
Additive OutSrc. & 0.858 & 0.886 \\
Additive OutSrc. + \OutAttenMul{} & 0.561 & 0.570  \\
Additive OutSrc. + \OutSoftMax{} & 0.525 & 0.531 \\
Additive OutSrc. + \OutAttenMul{} + \OutSoftMax{} & 0.174 & 0.216 \\\bottomrule
\end{tabular}
\label{t:ablation}
\end{table}

\begin{figure}[t!]
    \centering
    \includegraphics[width=\linewidth]{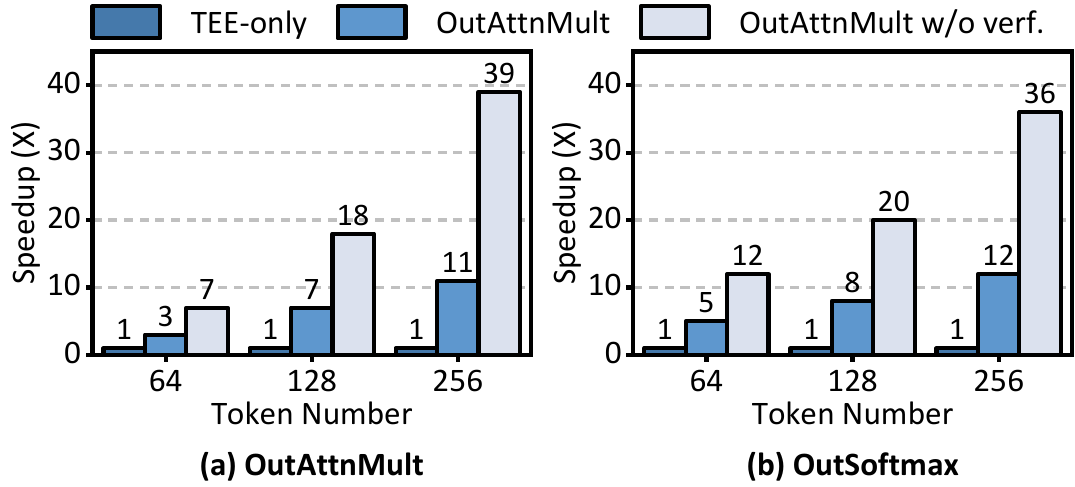}
    \caption{Performance comparison of \OutAttenMul{} (a) and \OutSoftMax{} (b) over TEE-only execution.}
    \label{fig:ablation}
\end{figure}

\noindent\textbf{Performance Benchmark on \OutAttenMul{} and \OutSoftMax{}. } 
To investigate the effectiveness of our proposed methods, \OutAttenMul{} and \OutSoftMax{}, across different matrix sizes and token counts, we conducted a performance benchmark. The results are illustrated in Figure~\ref{fig:ablation}. In Figure~\ref{fig:ablation} (a), we observe that \OutAttenMul{} secures progressively higher speedups with the growth in square matrix dimensions, from a \(3\times\) enhancement with a dimension of \(256\) to an \(11\times\) improvement with a dimension of \(1024\). This performance increase is attributed to \OutAttenMul{}'s efficiency in reducing the complexity of matrix multiplications from \(\mathcal{O}(n^3)\) to \(\mathcal{O}(n^2)\) within the TEEs. The limited memory capacity of the TEEs, which exacerbates computational overhead, further accentuates the performance disparity between \OutAttenMul{} and conventional TEE-only computation.

Figure~\ref{fig:ablation} (b) highlights the performance improvements introduced by \OutSoftMax{}. With matrix sizes expanding from \(256\) to \(1024\) dimensions, the observed speedups range from \(5\times\) to \(12\times\). These enhancements stem from the relative efficiency of multiplication over exponentiation, with \OutSoftMax{} effectively transforming exponential operations within the TEEs into multiplications. The observed trend of increased speedups with larger matrix dimensions can be linked to the TEE's constrained memory, necessitating more frequent paging for larger matrices and thereby reaping greater benefits from our optimization strategies.


\subsection{Analysis of Proposed Techniques}

\noindent\textbf{Breakdown analysis of \OutAttenMul{}.} In Table~\ref{t:breakdown_matmul}, we present a latency breakdown for each stage of the \OutAttenMul{} process, as applied to a single attention matrix multiplication of the BERT model. The results reveal that the offline phase within the TEEs is small, clocking in at $0.169$ ms. This is because our \OutAttenMul{} also outsources \(R_QR_K^T\) to the accelerator, compared to the scheme used in Slalom~\cite{tramer2018slalom} which requires precomputing matrix multiplications. In the online phase, the Embedded Additive Outsource requires slightly more time, at $0.152$ ms, due to the necessity for matrix permutation and a combination of two subtractions and an addition. The Recovery stage which needs scalar multiplication and addition requires $0.498$ ms. The Integrity Check stage, however, exhibits the highest latency within the TEEs, i.e., $0.763$ ms, necessitated by two vector-matrix multiplications to confirm the GPU's computation integrity. For the GPU, the matrix multiplication tasks in Embedded Additive Outsource stages are executed rapidly, leveraging the GPU's superior performance for matrix operations and contributing only $0.209$ ms.

Summing up, the total online latency for the TEEs is $1.413$ ms, while the GPU contributes an additional $0.209$ ms, leading to an aggregate end-to-end latency of $1.622$ ms for the complete \OutAttenMul{}.

\begin{table}[ht!]
\centering
\setlength{\tabcolsep}{9pt}
\caption{Latency breakdown of \OutAttenMul{}.}
\begin{tabular}{lcc}\toprule
Stage & TEE (ms) & GPU (ms) \\\midrule\midrule
Offline & 0.169 & -  \\
Embedded Additive Outsource & 0.152 & 0.209 \\
Recovery & 0.498 & - \\
Integrity Check & 0.763 & - \\\midrule
Online Total & 1.413 & 0.209 \\\bottomrule
\end{tabular}
\label{t:breakdown_matmul}
\end{table}

\noindent\textbf{Breakdown analysis of \OutSoftMax{}.} To investigate the latency various different stages in \OutSoftMax{}, we also conduct breakdown analysis in Table~\ref{t:breakdown_softmax} with token number of $256$. The offline phase within the TEEs is the most time-intensive, at $5.815$ ms, largely owing to the sampling of a random vector and computing the exponentials for each of its elements. During the Outsource Masked \(e^x\) stage, the TEEs handles element-wise multiplications of \(e^{x_i}\) and \(e^{r_i}\), contributing to the online latency. Similarly, the Division in the TEEs phase requires element-wise divisions, further adding to the TEE's workload. The Integrity Check, crucial for ensuring the correctness of the GPU's computations, involves additional element-wise multiplications by the TEEs to validate the results against \(e^{hashX}\). Overall, the online portion of the \OutSoftMax{} process necessitates $1.190$ ms from the TEEs and only $0.138$ ms from the GPUs. This breakdown analysis illustrates that speedup is attributed to converting the operation within TEEs from element-wise exponential to much cheaper element-wise multiplication.

\begin{table}[ht!]
\centering
\setlength{\tabcolsep}{14pt}
\caption{Latency breakdown of \OutSoftMax{}.}
\begin{tabular}{lcc}\toprule
Stage & TEE (ms) & GPU (ms) \\\midrule\midrule
Offline & 5.815 & -  \\
Outsource Masked $e^x$ & 0.327 & 0.138 \\
Division in TEE & 0.381 & - \\
Integrity Check & 0.482 & - \\\midrule
Online Total & 1.190 & 0.138 \\\bottomrule
\end{tabular}
\label{t:breakdown_softmax}
\end{table}

\noindent\textbf{Ablation on the value of random coefficient \(a\) in \UVerify{} for \OutSoftMax{}.} Our analysis of \OutSoftMax{} reveals the impact of the maximum value of random coefficients (\(a_{\text{max}}\)) on computational efficiency. In trials using a \(256 \times 256\) matrix, we observed that increasing \(a_{\text{max}}\) from 2 to 7 leads to a proportional decrease in speedup—from a maximum of \(7.4\times\) down to \(5.3\times\). This trend demonstrates a linear relationship between \(a_{\text{max}}\) and the speedup, which is consistent with our complexity analysis, confirming that the number of required multiplications is directly tied to \(a_{\text{max}}\). By carefully choosing an optimal \(a_{\text{max}}\), \OutSoftMax{} can effectively minimize the TEE's latency by simplifying \(\mathcal{O}(n)\) exponential operations to \(\mathcal{O}(n)\) multiplications.

\begin{figure}[ht!]
    \centering
    \centerline{\includegraphics[width=0.6\linewidth]{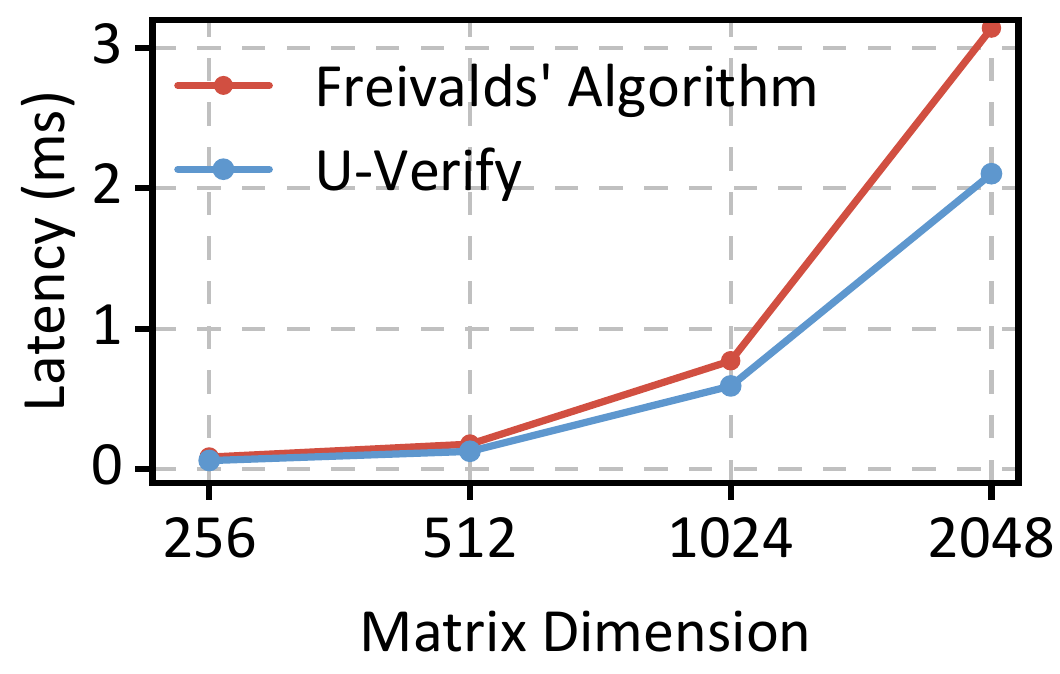}}
    \caption{Comparison between Freivalds' algorithm and \UVerify{} on linear attention multiplication. Note that Freivalds' algorithm is not applicable for non-linear operations such as \texttt{softmax}, unlike our \UVerify{} method. }
    \label{fig:uverify}
\end{figure}

\noindent\textbf{Comparison between Freivald's algorithm and \UVerify{}.} 
As Section~\ref{sec:method3} shows, it's important to highlight that Freivalds' algorithm cannot be used for verifying non-linear operations like \texttt{softmax}, in contrast to our \UVerify{} approach.
In assessing the verification efficiency on linear attention multiplication, we compared our \UVerify{} with the Freivalds' algorithm~\cite{freivalds1977probabilistic}. Figure~\ref{fig:uverify} indicates that \UVerify{} delivers an approximate \(33\%\) reduction in latency when compared to Freivalds' method. Freivalds' algorithm typically requires three vector-matrix multiplications to verify the product of \( A^{n\times n} \) and \( B^{n\times n} \) equals \( C^{n\times n} \), calculating \( Cs \), \( Bs \), and subsequently \( A(Bs) \). In contrast, \UVerify{} streamlines this process to just two vector-matrix multiplications. It achieves this by embedding a hash row, \( hashA \), within \( A \) and outsourcing \( hashA \cdot B \) to the GPU. Consequently, the TEEs merely need to verify that \( h_A \cdot C \) corresponds to \( hashA \cdot B \), leading to the observed efficiency gains.

\noindent\textbf{Accelerator Studies.} In addition to evaluating legacy GPUs, we also explored other accelerator platforms to assess the adaptability of our approach.

\noindent\textit{FPGAs.} To evaluate the versatility of \name { } with alternative accelerators, we conducted experiments using the Xilinx Alveo U280 FPGA. This FPGA features 8 GB of HBM2 memory, 32 GB of DDR memory with a bandwidth of 460 GBps, and 1,304K logic cells. It connects to the host machine via a PCIe Gen3x16 I/O interface. We utilized the Vitis BLAS Library for \OutAttenMul{} operations and the Vitis AI Library for outsourcing the \OutSoftMax{}. The results showed that outsourcing computations to the FPGA significantly reduced latency compared to performing all operations within a TEE. Specifically, \name { } achieved latency reductions ranging from $1.93\times$ to $3.25\times$, demonstrating the effectiveness of our outsourcing strategy across different accelerator platforms.

\noindent\textit{TPUs.} We also tested the adaptability of \name { } on Google's TPU VM v3-8, which is equipped with 128 GB of total memory distributed across 8 chips and offers an interconnect bandwidth of 100 GBps. For implementation, we used TensorFlow 2.1. The results revealed that leveraging a TPU for outsourcing operations drastically reduced latency compared to relying exclusively on a TEE for computations. Notably, \name { } achieved latency reductions between $7.43\times$ and $10.91\times$ when utilizing the TPU, highlighting the significant advantages of our outsourcing strategy across diverse computational platforms. 

\section{Conclusion}
In this paper, we propose \name { }, an innovative framework to safeguard the privacy and integrity of Transformer inference services in the cloud setting. We design secure and efficient modules, \OutAttenMul{} and \OutSoftMax{}, to outsource bottleneck computations in Transformers including the non-additive attention multiplication and non-linear \texttt{softmax} functions. We further propose a novel verification scheme \UVerify{} to ensure the integrity of the outsourced computation. Extensive experiments show that \name { } offers from $4.0\times$ to $6.1\times$ performance improvement over prior works for private verifiable inference without sacrificing accuracy.

\bibliographystyle{IEEEtran}
\bibliography{sections/example_paper,lou}

\appendix

\section{Feasible Set Construction}

\label{app:feasible_set}
In this section, we will outline the method to construct the feasible set, $F(\hat{X})$ for a given transformed matrix $\hat{X}$ (the $X$ could be $Q$ or $K^T$).

We refer to the $r_i$ in the random vectors $R=[r_1,...,r_{t-n}]$ as mask vectors. Additionally, let $\subset_{R}$ represent a uniform random sampling. The idea is to back-trace and compute the set of possible of pre-images. Now the feasible set is constructed as follows:
\begin{enumerate}
    \item Select the set of $t-n$ indices uniformly at random:
        \begin{equation}
            \Omega\subset_R[m], |\Omega|=t-n
        \end{equation}
    $\Omega$ represents a possible set of indices that correspond to the mask vectors.
    \item The corresponding set of mask vectors is:
        \begin{equation}
            \Phi_\Omega=\{\hat{x_i}|i\in\Omega\}
        \end{equation}
    \item Let $\overline{\Phi}_\Omega=\{\hat{x_i}|i\in[n] \backslash \Omega\}$ be the set of transformed original vectors. Additionally, let $\overline{X}=[\overline{x}_1,...,\overline{x}_n]$ where $\overline{x}_i\in\overline{\Phi}_\Omega$ and $\overline{x}_i\neq\overline{x}_j$, $i,j\in[n]$, $i\neq j$.
    \item Sample a random permutation $\sigma$ to recover the shuffled matrix. Thus $\overline{X}_\alpha=[\overline{x}_{\sigma(1)},...,\overline{x}_{\sigma(n)}]$ represents a possible transformed matrix. 
    \item Compute
        \begin{equation}
        \footnotesize
            \begin{gathered}
            \forall i \in [n] \\
            \mathcal{F}_{\Omega,\sigma}^i(\hat{X})=\{x|x=d\cdot(\overline{x}_{\sigma(i)}-\hat{x}'),d\in\mathbb{R},\hat{x}'\subset_R\Phi_\Omega\}
            \end{gathered}
        \end{equation}
    $\mathcal{F}_{\Omega,\sigma}^i(\hat{X})$ denotes the set of possible values for the vector $x_i$ for the give $\Omega$ and $\sigma$.
    \item Clearly, we have
        \begin{equation}
            \mathcal{F}(\hat{X})=\bigcup_{\Omega}\bigcup_{\sigma}\mathcal{F}_{\Omega,\sigma}(\hat{X})
        \end{equation}
\end{enumerate}
Clearly, larger the value of obfuscation ratio $r$, greater is the size of $\Omega$ and consequently, $\mathcal{F}(\hat{X})$. Additionlly, it is evident that $\mathcal{F}(\hat{X_a})\supset \mathcal{F}(\hat{X_b})$ where $t_a=|\hat{X_a}|>|\hat{X_b}|=t_b$ (equivalently, $t_a>t_b$).

\section{Security Analysis of \OutAttenMul{}}
\label{app:security_attenmul}

For the transformed $\widetilde{Q}$ and $\widetilde{K^T}$, let $F(\widetilde{Q})$ and $F(\widetilde{K^T})$ represent the set of original matrixs that could have been transformed to $\widetilde{Q}$ and $\widetilde{K^T}$, i.e., the set of possible pre-images for $\widetilde{Q}$ and $\widetilde{K^T}$. We call them \textit{feasible set} for $\widetilde{Q}$ and $\widetilde{K^T}$. The construction of \textit{feasible set} can be found in appendix~\ref{app:feasible_set}.

\noindent \textbf{Theorem 1.} For a $\widetilde{QK^T}$ computation and a given view of the GPU $\text{View}_{\text{GPU}}=(\widetilde{Q}, \widetilde{K^T}, \widetilde{QK^T}, (QK^T)')$, where \((QK^T)'\) is used to operate subsequent computation in GPU, has been transformed by TEE. We have:
\begin{equation}
\label{e:Q}
    \begin{gathered}
    \forall (Q_i,Q_j) \in \mathcal{F}(\widetilde{Q})\times \mathcal{F}(\widetilde{Q}) \\
    \text{Pr}[Q=Q_i|\text{View}_{\text{GPU}}]=\text{Pr}[Q=Q_j|\text{View}_{\text{GPU}}]
    \end{gathered}
\end{equation}
and
\begin{equation}
\label{e:K}
    \begin{gathered}
    \forall (K^T_i,K^T_j) \in \mathcal{F}(\widetilde{K^T})\times \mathcal{F}(\widetilde{K^T}) \\
    \text{Pr}[K^T=K^T_i|\text{View}_{\text{GPU}}]=\text{Pr}[K^T=K^T_j|\text{View}_{\text{GPU}}]
    \end{gathered}
\end{equation}

The above theorem states that, on observing transformed input matrices $\widetilde{Q}$ and $\widetilde{K^T}$, output matrix $\widetilde{QK^T}$ and next input matrix $(QK^T)'$, an attacker cannot distinguish between two matrices that belong to its feasible set. Thus, the feasible sets act as cloaking regions for the original matrices.

\noindent \textit{Proof.} First, we present two helper lemmas as follows.

\noindent \textbf{Lemma 1.} \textit{Attacker cannot reconstruct $QK^T$ from $\widetilde{QK^T}$ and $(QK^T)'$.} Recall that the attacker (GPU) computes $\widetilde{QK^T}=\widetilde{Q}\cdot\widetilde{K^T}$ and subsequently transmits it to TEE. Within the TEE, $\widetilde{QK^T}$ undergoes a recovery process to $QK^T$ through a specific linear transformation, denoted as $f(\cdot)$. Following this, $QK^T$ is subjected to a masking process via another linear transformation, represented as $g(\cdot)$, resulting in the matrix $(QK^T)'$, which is then relayed back to the GPU for subsequent computations. Given the absence of knowledge regarding the intricacies and parameters of the linear transformations $f(\cdot)$ and $g(\cdot)$ on the part of the attacker, it is postulated that the reconstruction of the intermediate matrix $QK^T$ from the accessible matrices $\widetilde{QK^T}$ and $(QK^T)'$ is not computationally feasible for the adversarial entity.

\noindent \textbf{Lemma 2.} \textit{Random masked vectors \(r_i \in R_Q\) or \(r_j \in R_K\) are indistinguishable from transformed vectors $q_i \in Q$ or $k_j \in K$ which are the same shape.} Note that $\widetilde{Q}$ and $\widetilde{K^T}$ are embedded in a field $\mathbb{F}$. Thus clearly, masking the inputs is equivalent to applying a one-time pad~\cite{katz2007introduction}. And the \(\widetilde{Q}\) and \(\widetilde{K^T}\) are permuted before sending to GPU.


\noindent \textbf{Theorem 2.} Given the intricacies of the transformation process applied to the $n$-dimension matrix $X$, we can quantify the probability of correctly deducing the original matrix from its transformed version by the attacker. This quantification is simplified under certain assumptions, notably regarding the scalar coefficients used in the transformation. The theorem assumes these scalar coefficients are integers constrained within the range $(-L,L)$. This simplification is a significant factor in calculating the overall probability and is chosen for its practicality in mathematical modeling and computation.

\begin{equation}
\text{Pr[Correct]} = \frac{\binom{n}{n}}{\binom{2n}{n}} \times \left(\frac{1}{2n-n}\right)^n \times \left(\frac{1}{2L}\right)^n \times \frac{1}{n!}
\label{e:verify_outmul}
\end{equation}

This equation encapsulates the aggregate probability, taking into account various factors including the selection of the correct vectors, identification of corresponding masks, accurate prediction of scalar coefficients, and the reordering of the shuffled matrix.

\noindent \textit{Proof.}
The proof of Theorem 2 integrates the factors influencing the overall probability of an attacker correctly guessing the original matrix $X$. The probability is derived from the combination of several factors:
\begin{enumerate}
    \item Selection of Original Vectors: The correct identification of the original $n$ vectors from a total of $2n$ vectors is represented by the combinatorial ratio $\frac{\binom{n}{n}}{\binom{2n}{n}}$, indicating the statistical likelihood of choosing the exact subset of original vectors.
    \item Mask Identification: In the obfuscation process, correctly identifying the appropriate masks from the remaining $n$ vectors has a probability of $\left(\frac{1}{2n-n}\right)^n$, assuming each choice is independent and uniform.
    \item Scalar Coefficients: The scalar coefficients, which are integers within the range $(-L,L)$, have a uniform probability distribution. Thus, the probability of correctly guessing all coefficients is $\left(\frac{1}{2L}\right)^n$.
    \item Order Recovery: The likelihood of correctly re-establishing the original sequence in the shuffled matrix is represented by the permutation probability $\frac{1}{n!}$.
\end{enumerate}

\section{Security Analysis of \OutSoftMax{}}

Given the transformed vector \(X' = [x_1-r_1, ..., x_i-r_i]\), it is infeasible for an attacker to recover the original vector \(X = [x_1, x_2, ..., x_i]\). The level of security is guaranteed by the additive secret sharing~\cite{cramer2015ass, demmler2015ass2} which assumes that the attacker lacks knowledge of the random vector \([r_1, r_2, ..., r_i]\), which are essential for the reconstruction of the original vector.

\section{Security Analysis of \UVerify{}}
\subsection{Verification of \OutSoftMax{}}
\label{app:security_uverify_softmax}
Given the vector share of the GPU \([x_1-r_1,...,hashX-r_{hash},...,x_n-r_n]\), GPU is expected to return \([e^{x_1-r_1},...,e^{hashX-r_{hash}},...,e^{x_n-r_n}]\). Here we consider three specific types of attack as the supplement:

(a) the attacker tampers the results to 
\begin{equation}
    [e^{x_1-r_1+\Delta_1},...,e^{hashX-r_{hash}},...,e^{x_n-r_n}]
\end{equation}

(b) the attacker tampers the results to 
\begin{equation}
    [e^{x_1-r_1}+\Delta_1,...,e^{hashX-r_{hash}},...,e^{x_n-r_n}]
\end{equation}

(c) the attacker swap the positions of different elements.

According to the verification process in Equation~\ref{e:hash_softmax}, the TEE will check if:

\begin{equation}
    \left(e^{x_1+\Delta_1}\right)^{a_1}\prod_{i=2}^{n} (e^{x_i})^{a_i} = e^{hashX}
\end{equation}

For the attack (a), the attacker needs also change:
\begin{equation}
    e^{hashX-r_{hash}} \rightarrow e^{hashX-r_{hash}}e^{\Delta_1a^1}
\end{equation}

So it is crucial to discern the index of \(hashX\) as well as the coefficients \(a_i\) that the TEE uses for verification of \(x_i\) in Equations~\ref{e:hash_softmax}.

\noindent\textbf{Lemma 3.} \textit{Given the vector \([x_1-r_1, ..., hashX - r_{hash},...,x_n-r_n]\), \(hashX - r_{hash}\) is indistinguishable from any transformed \(x_i-r_i\).} Since \(x_i-r_i\) and \(hashX - r_{hash}\) are embedded within a field \(\mathbb{F}\) and the \(hashX - r_{hash}\) is inserted by the TEE in an obfuscated location.

\noindent \textbf{Theorem 3.} Building upon Lemma 3, we can determine the probability of an attacker successfully tampering with \(k\) bits using attack (a) of an \(n\)-dimensional vector without detection by \(\text{\UVerify{}}\):
\begin{equation}
    \text{Pr[Correct]} = \frac{1}{n} \times \left(\frac{1}{2L}\right)^k
\end{equation}
The factor \(\frac{1}{n}\) reflects the probability of accurately identifying \(hashX\), echoing the premise of Lemma 3. The following term, \(\left(\frac{1}{2L}\right)^k\), represents the probability of correctly guessing all \(k\) coefficients \(a_i\), where these coefficients are integers sampled from the interval \((-L, L)\).

For attack (b), the tampered item \(e^{x_1-r_1}+\Delta_1\) will first be multiplied by \(e^{r_1}\) during the verification in TEE, and then the TEE will check if:

\begin{equation}
    \left(e^{x_1}+\Delta_1e^{r_1}\right)^{a_1}\prod_{i=2}^{n} (e^{x_i})^{a_i} = e^{hashX}
\end{equation}

so the attacker needs also know the \(r_1\), \(a_1\) and \(e^{x_1}\) to successful attack. The success probability is:

\begin{equation}
    \text{Pr[Correct]} = \frac{1}{n} \times \left(\frac{1}{2L}\right)^k \times \left(\frac{1}{2L}\right)^k \times \left(\frac{1}{2L}\right)^k
\end{equation}

For attacker (c), because the secret vector \(\boldsymbol{a}\) will used in the integrity check as Equation~\ref{e:hash_softmax}, any mismatch of \(a_i\) and \(x_i\) will leads to the failure verification in Equation~\ref{e:hash_softmax}.

\subsection{Verification of \OutAttenMul{}} 
\label{app:security_uverify_atten}
In the \UVerify{} process, the TEE first generates \(hashQ\) by \(hash_Q=h_Q\cdot Q\), and then outsources the concatenated matrix by \OutAttenMul{}. The security are guaranteed by the lower probability of identify the \(hashQ\) from the \(\widetilde{Q}\) quantized in Equation~\ref{e:verification_attention}. Upon the assumption that the attacker cannot identify the \(hashQ\), the integrity are guaranteed using the variants of an algorithm by Freivalds~\cite{freivalds1977probabilistic}. 

\noindent\textbf{Theorem 4.} (Freivalds) \textit{Let \(Q\), \(K^T\) and \(Z\) be \(n\times n\) matrices on over a field \(\mathbb{F}\) and let \(s\) be a uniformly random vector in \(\mathbb{S}\subseteq\mathbb{F}\). Then \(\text{Pr}[sZ=(sQ)K^T|Z\neq QK^T]=\text{Pr}[s(Z-QK^T)=0|(Z-QK^T)\neq 0]\leq \frac{1}{|\mathbb{S}|}\)}.

\label{s:security_uverify}

\section{Neural Network Details}
\label{app:experiments}

\begin{table}[ht!]
\centering
\vspace{-0.15in}
\setlength{\tabcolsep}{4pt}
\caption{Accuracy of models and quantized models.}
\begin{tabular}{lccccc}\toprule
 & Layers & Parameters & Metrics & Original & Quantized \\\midrule
ViT-16B & $12$ & $87$M & ACC \(\uparrow\) & $74.6$\% & $72.7\%$ \\
CLIP & $24$ & $151$M & ACC \(\uparrow\) & $73.6$\% & $73.3\%$ \\
BERT & $12$ & $110$M & ACC \(\uparrow\) & $92.4$\% & $91\%$ \\
LLaMA & $32$ & $7$B & Perplexity \(\downarrow\) & $5.68$ & $5.89$ \\\bottomrule
\end{tabular}
\label{t:quantize}
\end{table}
In Table~\ref{t:quantize}, we report the performance of evaluated models with and without the simple quantization scheme described in Section~\ref{sec:setup}. Quantization results in at most a 1.9\% drop in accuracy and a 0.21 perplexity increase.

\end{document}